\documentclass[12pt,english,aps, final]{revtex4}
\usepackage[T1]{fontenc}
\usepackage[latin9]{inputenc}
\usepackage{amsmath}
\usepackage{cancel}
\usepackage{stackrel}
\usepackage{graphicx}
\usepackage{esint}
\usepackage{amssymb}

\makeatletter

\providecommand{\tabularnewline}{\\}

\@ifundefined{textcolor}{}
{%
 \definecolor{BLACK}{gray}{0}
 \definecolor{WHITE}{gray}{1}
 \definecolor{RED}{rgb}{1,0,0}
 \definecolor{GREEN}{rgb}{0,1,0}
 \definecolor{BLUE}{rgb}{0,0,1}
 \definecolor{CYAN}{cmyk}{1,0,0,0}
 \definecolor{MAGENTA}{cmyk}{0,1,0,0}
 \definecolor{YELLOW}{cmyk}{0,0,1,0}
}

\makeatother

\usepackage{babel}
\begin{document}

\title{Dispersion Approach in Two-Loop Calculations}

\author{A. Aleksejevs}

\affiliation{Grenfell Campus of Memorial University, Canada }
\begin{abstract}
The higher-order corrections become increasingly important with experiments
reaching sub-percent level of uncertainty as they look for physics
beyond the Standard Model. Our goal is to address the full set of
two-loop electroweak corrections to M{\o}ller or electron-proton scattering.
It is a demanding task which requires an application of various approaches
where two-loop calculations can be automatized. We choose to employ
dispersive sub-loop insertion approach and develop two-loop integrals
using two-point functions basis. In that basis, we introduce a partial
tensor reduction for many-point Passarino-Veltman functions, which
later could be used in computer algebra packages. In this paper, we
have considered self-energy, triangle and box sub-loop insertions into
self-energy, vertex and box topology.
\end{abstract}
\maketitle
\tableofcontents{}

\section{Introduction}

The electroweak precision searches for the physics beyond the Standard
Model (BSM) frequently demand a sub-percent level of accuracy from
both experiment and theory. From the theory perspective, this can
be achieved by extending the perturbation expansion of the scattering
matrix element to the two-loop level. However, since the electroweak
(EW) interaction usually introduce different-mass propagators and
higher-order tensor Feynman integrals, the two-loop EW calculations
can become increasingly complicated. In general, in electroweak case, 
it is very challenging or even not possible to find analytic results 
beyond the one-loop level, so one would either have to resort to various 
approximations or purely numerical methods.  See, for example, a 
comprehensive overview of a variety of numerical loop integration 
techniques in \cite{Freitas1}, a general case of the two-loop two-point
 function for arbitrary masses in \cite{Kreimer}, and a method of calculating 
 scalar propagator and vertex functions based on a double integral 
 representation in \cite{Czarnecki} and \cite{Frink}. The more recent 
 developments on analytical evaluation of two-loop self-energies can 
 be found in \cite{Adams1, Adams2, Adams3, Remiddi1, Bloch1, Bloch2}, 
 and on numerical evaluation of general n-point two-loop integrals 
 using sector decomposition in \cite{Borowka1, Borowka2}.  In addition, 
 we normally have to evaluate several thousands Feynman graphs. Obviously, 
 such a voluminous task should be delegated to the computer-based calculations. 
 Although there has been a strong progress, we still have an ongoing problem
of having to deal with cumbersome expressions and consequently be
forced to introduce approximations. In many existing techniques, the
tensor integrals have to be reduced to master scalar integrals which
increases size of the final expressions dramatically. In order to
address this, we employ a dispersive approach to sub-loop diagrams,
and introduce a partial tensor reduction of the two-loop graphs. In
general, a sub-loop can be represented through a dispersion tensor
integral operator with a relatively simple propagator-like structure.
Dispersion tensor integral numerator could be then absorbed into the
effective Feynman vertices or propagators, and the second-loop integration
will acquire an additional propagator. The idea of the sub-loop insertions
with the help of the dispersive approach was implemented for the self-energies
\cite{Bohm}, \cite{Hollik-1} and partially for the vertex graphs
with the help of Feynman parametrization \cite{Hollik-2}. We extend
this for self-energy, vertex and box sub-loop insertions of the general
tensor structure. In addition, we apply the reduction of the three-,
four-, and five-point tensor coefficient functions insertions to the
derivative representation of the two-point tensor function. Of course,
as in previous works, the treatment of the UV and IR divergences
requires subtractions derived from the two-loop EW counterterms and
introduction of the photon mass regulator. In ``Sub-Loop Approach'' section, 
we start with basic definitions and ideas of dispersive treatment of sub-loop insertions. After that, 
we consider self-energies, vertex and box insertions and provide a partial
tensor reduction. The section ``Numerical Example'' considers specific
examples of two-loop self energies with vertex-type insertions and
provide numerical comparison with \cite{Bohm} for the kinematical
region below and above threshold.

\section{Sub-loop Approach}

Generally, a dispersion relation allows to express a loop integral
through the known imaginary part: 
\begin{align}
L(q^{2})=\frac{1}{\pi}\stackrel[s_{0}]{\infty}{\int} & ds\frac{\Im L(s)}{s-q^{2}-i\epsilon}.\label{eq:1}
\end{align}
Here, $q^{2}$ is the external momentum squared and $s_{0}$ is the
branch point position on the real axis. The imaginary part $\Im L(q^{2})$
can be calculated from discontinuities of the loop integral using
Cutkosky rules. If we consider the sub-loop insertion represented
by self energy, triangle or box, we can extract an imaginary part
of two-, three-, and four-point tensor coefficient functions from
the routines such as FF \cite{FF} and LoopTools \cite{LoopTools},
which are already implemented numerically. This leaves us with a problem
of expressing the two-loop tensor integrals in terms of many-point
tensor coefficient function. We start with definition of a general
two-loop tensor integral in the dimensional regularization:
\begin{align}
I_{\mu_{1}...\mu_{G},\nu_{1}...\nu_{R}}^{N,M,P} & =\frac{\mu^{2(4-D)}}{(i\pi^{D/2})(i\pi^{D/2})}\int d^{D}q_{1}d^{D}q_{2}\cdot\nonumber \\
\label{eq:2}\\
 & \frac{q_{1,\mu1}...q_{1,\mu_{G}}q_{2,\nu_{1}}...q_{2,\nu_{R}}}{\stackrel[i=0]{N}{\Pi}[(q_{1}+k_{i,N})^{2}-m_{i,N}^{2}]\stackrel[j=0]{M}{\Pi}[(q_{2}+k_{j,M})^{2}-m_{j,M}^{2}]\stackrel[l=0]{P}{\Pi}[(q_{1}+q_{2}+k_{l,P})^{2}-m_{l,P}^{2}]},\nonumber 
\end{align}
where $q_{1,2}$ are the integration momenta in the first and second
loops, respectively. The momenta $k_{i,j,l}$ represent various combinations
of the external momenta $p_{i,j,l}$ from a two-loop graph. The masses
of internal particles are defined as $m_{i,j,l}$. For the processes
specifically related to the parity-violating scattering, a sub-loop
topology would be defined by either self-energy, triangle or box insertions.

\subsection{Self-Energy Sub-Loop}

\begin{figure}
\begin{centering}
\includegraphics[scale=0.45]{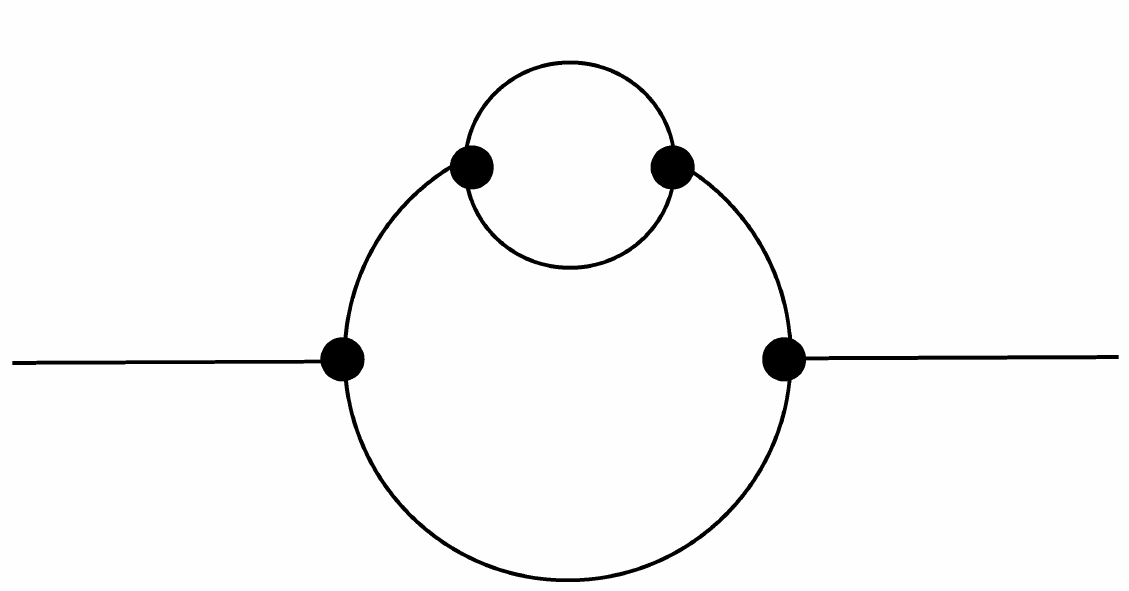}\includegraphics[scale=0.45]{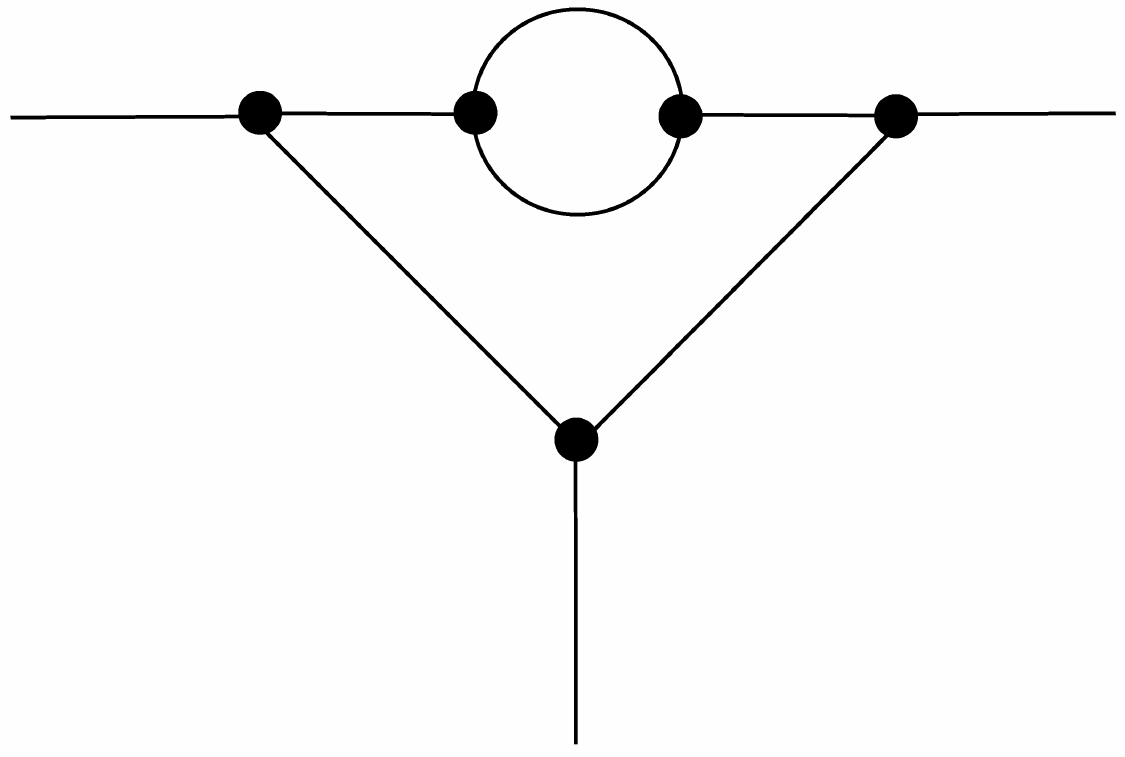}\includegraphics[scale=0.45]{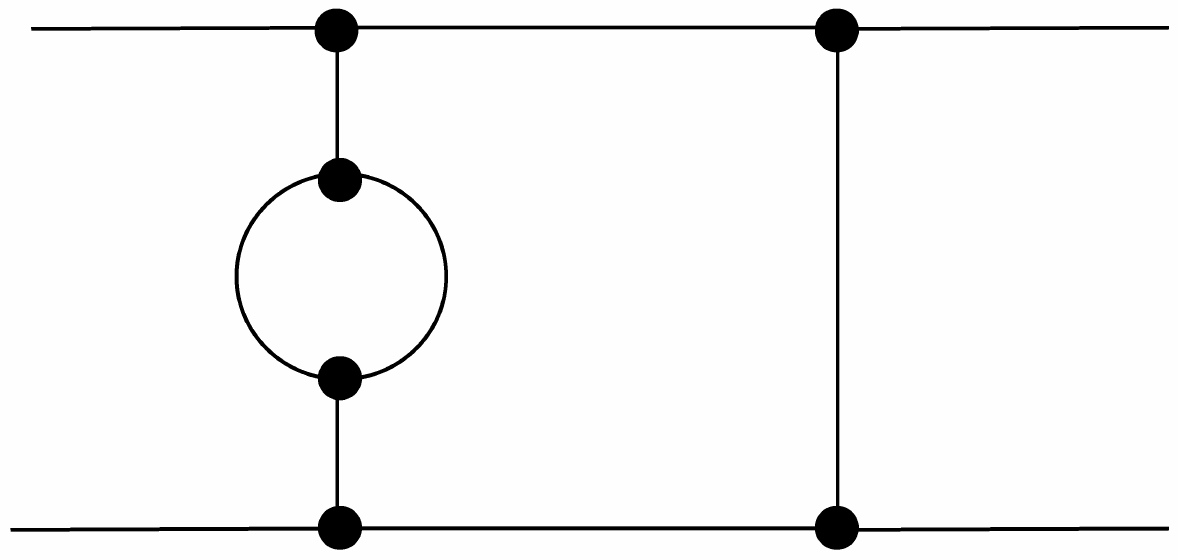}
\par\end{centering}
\caption{Examples of self-energy sub-loops in the self-energy, triangle and
box topologies. In general, self-energy could be applied to any internal
line.}

\label{f1}
\end{figure}

\begin{figure}
\begin{centering}
\includegraphics[scale=0.45]{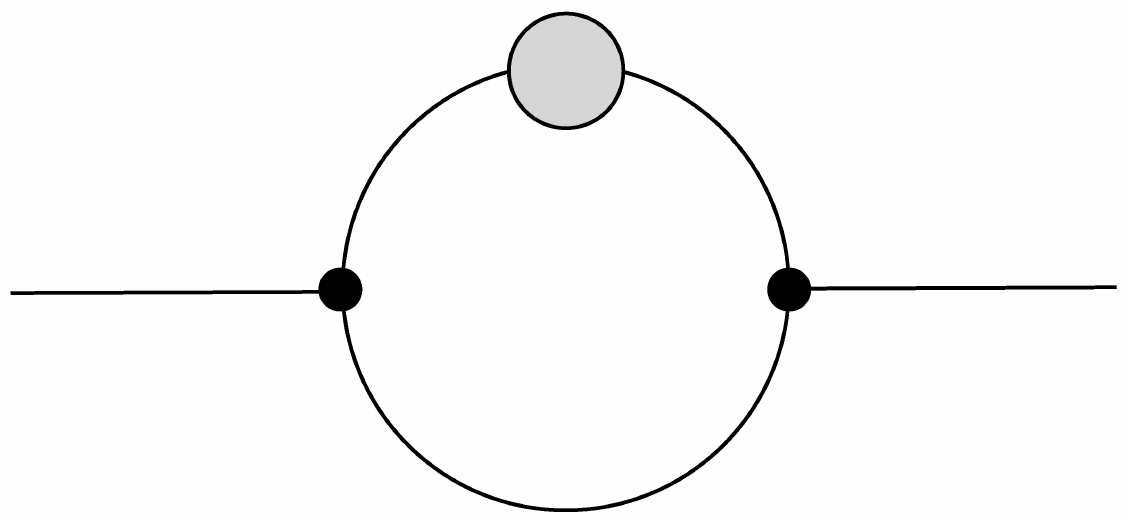}\includegraphics[scale=0.45]{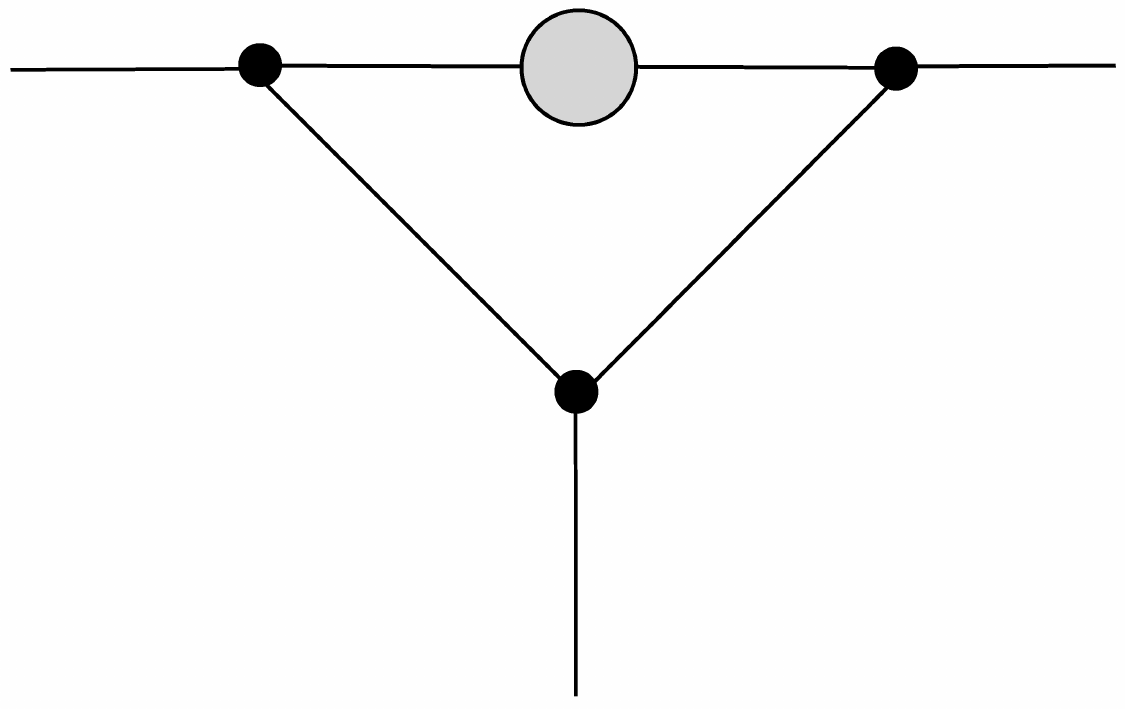}\includegraphics[scale=0.45]{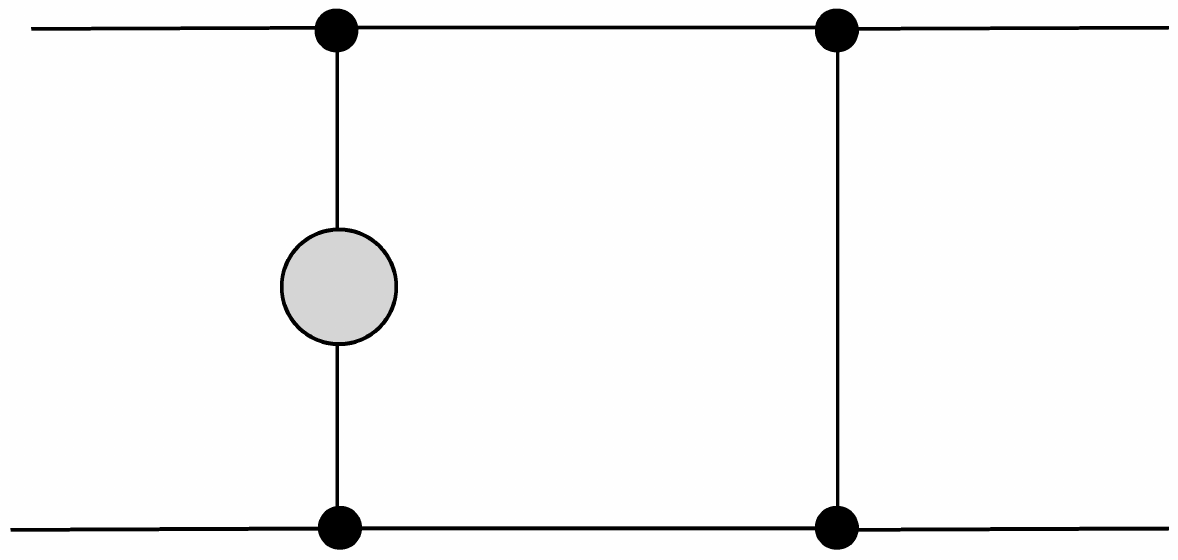}
\par\end{centering}
\caption{Reduced two-loop topologies after sub-loop self-energy insertions. }

\label{f2}
\end{figure}

The self-energy sub-loop could be inserted into another self-energy,
triangle or box topology (see Fig.(\ref{f1})).
After replacing self-energy sub-loop by the dispersion integral, graphs
from Fig.(\ref{f1}) could be reduced to graphs shown on the Fig.(\ref{f2}).
More specifically, for fermions or vector bosons, the self-energy sub-loop
can be defined in form of the Lorentz covariant terms: 
\begin{align}
 & \Sigma_{\mu\nu}^{V-V}(q)=\left(g_{\mu\nu}-\frac{q_{\mu}q_{\nu}}{q^{2}}\right)\Sigma_{T}^{V-V}\left(q^{2}\right)+\frac{q_{\mu}q_{\nu}}{q^{2}}\Sigma_{L}^{V-V}\left(q^{2}\right)\label{eq:3}\\
\nonumber \\
 & \Sigma^{f}(q)=\cancel{q}\omega_{-}\Sigma_{L}^{f}\left(q^{2}\right)+\cancel{q}\omega_{+}\Sigma_{R}^{f}\left(q^{2}\right)+m_{f}\Sigma_{S}^{f}\left(q^{2}\right).\label{eq:4}
\end{align}
Here, in Eq.(\ref{eq:3}), $\Sigma_{T,L}^{V-V}\left(q^{2}\right)$
represents transverse and longitudinal parts of diagonal and mixed vector boson 
self-energies. In Eq.(\ref{eq:4}), $\Sigma_{L,R,S}^{f}\left(q^{2}\right)$
represents left, right and scalar parts of the fermion self-energy
graph. The $\omega_{\pm}=\frac{1\pm\gamma_{5}}{2}$ are usual left/right
chirality projectors. Each of the blocks $\Sigma$ in Eqs.(\ref{eq:3})
and (\ref{eq:4}) can be written in terms of Passarino-Veltman two-point
tensor coefficient functions. Then, each of the two-point tensor coefficient
functions $B_{i,ij,ijk}\left(q^{2},m_{\alpha}^{2},m_{\beta}^{2}\right)$
can be replaced by the dispersion integral:
\begin{align}
 & B_{i,ij,ijk}\left(q^{2},m_{\alpha}^{2},m_{\beta}^{2}\right)=\frac{1}{\pi}\stackrel[\left(m_{\alpha}+m_{\beta}\right)^{2}]{\infty}{\int}ds\frac{\Im B_{i,ij,ijk}\left(s,m_{\alpha}^{2},m_{\beta}^{2}\right)}{s-q^{2}-i\epsilon},\label{eq:5}
\end{align}
where $\Im B_{i,ij,ijk}\left(s,m_{\alpha}^{2},m_{\beta}^{2}\right)$
can be easily computed using LoopTools or FF libraries. As a result,
Eqs.(\ref{eq:3}) and (\ref{eq:4}) can be re-written in the following
form:
\begin{align}
 & \Sigma_{\mu\nu}^{V-V}(q)=\frac{1}{\pi}\stackrel[\alpha,\beta]{}{\varSigma}\stackrel[\left(m_{\alpha}+m_{\beta}\right)^{2}]{\infty}{\int}ds\frac{1}{s-q^{2}-i\epsilon}\left[\left(g_{\mu\nu}-\frac{q_{\mu}q_{\nu}}{q^{2}}\right)\Im\Sigma_{T}^{V-V}\left(s,m_{\alpha}^{2},m_{\beta}^{2}\right)+\frac{q_{\mu}q_{\nu}}{q^{2}}\Im\Sigma_{L}^{V-V}\left(s,m_{\alpha}^{2},m_{\beta}^{2}\right)\right]\nonumber \\
\label{eq:6}\\
 & \Sigma^{f}(q)=\frac{1}{\pi}\stackrel[\alpha,\beta]{}{\varSigma}\stackrel[\left(m_{\alpha}+m_{\beta}\right)^{2}]{\infty}{\int}ds\frac{1}{s-q^{2}-i\epsilon}\left[\cancel{q}\omega_{-}\Im\Sigma_{L}^{f}\left(s,m_{\alpha}^{2},m_{\beta}^{2}\right)+\cancel{q}\omega_{+}\Im\Sigma_{R}^{f}\left(s,m_{\alpha}^{2},m_{\beta}^{2}\right)+m_{f}\Im\Sigma_{S}^{f}\left(s,m_{\alpha}^{2},m_{\beta}^{2}\right)\right].\nonumber 
\end{align}
The summation in Eq.(\ref{eq:6}) is done over all possible internal
particles in the self-energy sub-loop carrying masses $m_{\alpha}$
and $m_{\beta}$. Using Eq.(\ref{eq:6}), we can now write general
expressions for the two-loop topologies in Fig.(\ref{f1}). In a case
of a sub-loop represented by a vector boson self-energy, we get: 
\begin{align}
I_{\mu_{1}\mu_{2},\nu_{1}...\nu_{R}}^{1,M,1} & =\frac{1}{\pi}\frac{\mu^{(4-D)}}{(i\pi^{D/2})}\stackrel[\alpha,\beta]{}{\varSigma}\stackrel[\left(m_{\alpha}+m_{\beta}\right)^{2}]{\infty}{\int}ds\int d^{D}q_{2}\cdot\nonumber \\
\label{eq:7}\\
 & \frac{q_{2,\nu_{1}}...q_{2,\nu_{R}}}{(s-q_{2}^{2}-i\epsilon)\stackrel[j=0]{M}{\Pi}[(q_{2}+k_{j,M})^{2}-m_{j,M}^{2}]}F_{\mu_{1}\mu_{2}}\left(q_{2},s,m_{\alpha},m_{\beta}\right),\nonumber 
\end{align}
where $F_{\mu_{1}\mu_{2}}\left(q_{2},s,m_{\alpha},m_{\beta}\right)$
is defined as
\begin{align}
 & F_{\mu_{1}\mu_{2}}\left(q_{2},s,m_{\alpha},m_{\beta}\right)=\left(g_{\mu_{1}\mu_{2}}-\frac{q_{2\mu_{1}}q_{2\mu_{2}}}{q_{2}^{2}}\right)\Im\Sigma_{T}^{V-V}\left(s,m_{\alpha}^{2},m_{\beta}^{2}\right)+\frac{q_{2\mu_{1}}q_{2\mu_{2}}}{q_{2}^{2}}\Im\Sigma_{L}^{V-V}\left(s,m_{\alpha}^{2},m_{\beta}^{2}\right).\label{eq:8}
\end{align}
For the sub-loop insertion stemming from fermion self-energy we can
write:
\begin{align}
I_{\nu_{1}...\nu_{R}}^{1,M,1} & =\frac{1}{\pi}\frac{\mu^{(4-D)}}{(i\pi^{D/2})}\stackrel[\alpha,\beta]{}{\varSigma}\stackrel[\left(m_{\alpha}+m_{\beta}\right)^{2}]{\infty}{\int}ds\int d^{D}q_{2}\cdot\nonumber \\
\label{eq:9}\\
 & \frac{q_{2,\nu_{1}}...q_{2,\nu_{R}}}{(s-q_{2}^{2}-i\epsilon)\stackrel[j=0]{M}{\Pi}[(q_{2}+k_{j,M})^{2}-m_{j,M}^{2}]}G\left(q_{2},s,m_{\alpha},m_{\beta}\right),\nonumber 
\end{align}
and 
\begin{align}
 & G\left(q_{2},s,m_{\alpha},m_{\beta}\right)=\left[\cancel{q}_{2}\omega_{-}\Im\Sigma_{L}^{f}\left(s,m_{\alpha}^{2},m_{\beta}^{2}\right)+\cancel{q}_{2}\omega_{+}\Im\Sigma_{R}^{f}\left(s,m_{\alpha}^{2},m_{\beta}^{2}\right)+m_{f}\Im\Sigma_{S}^{f}\left(s,m_{\alpha}^{2},m_{\beta}^{2}\right)\right].\label{eq:10}
\end{align}
In both Eq.(\ref{eq:7}) and Eq.(\ref{eq:9}), the integration over
the second loop momenta can be written in the form of the Passarino-Veltman
many-point tensor coefficient functions. For self-energy sub-loop insertion specifically, we can 
simplify the second-loop integral by means of partial fraction decomposition. Since the momentum running  in the adjacent propagators and dispersion term $(s-q_2-i\epsilon)^{-1}$ is the same, we can write: 

\begin{align}
&\frac{1}{q^2_2-m_i^2}\cdot \frac{1}{q^2_2-s}\cdot \frac{1}{q^2_2-m_j^2}=
\nonumber\\
\nonumber\\
&\frac{1}{m_i^2-m_j^2}\cdot \left( \frac{1}{m_i^2-s} \cdot \frac{1}{q^2_2-m_i^2}  -  \frac{1}{m_j^2-s}\cdot  \frac{1}{q^2_2-m_j^2} \right)+\frac{1}{m_i^2-s} \cdot \frac{1}{m_j^2-s}\cdot \frac{1}{q^2_2-s}
\label{eq:10a} 
\end{align}

 This gives the following
result for a case of the vector boson self-energy insertion:
\begin{align}
I_{\mu_{1}\mu_{2},\nu_{1}...\nu_{R}}^{1,M,1} & =\frac{1}{\pi}\stackrel[\alpha,\beta]{}{\varSigma}\stackrel[\left(m_{\alpha}+m_{\beta}\right)^{2}]{\infty}{\int}ds\cdot\Bigg[L_{a,\mu_{1}\mu_{2},\nu_{1}...\nu_{R}}^{1,M,1}\left(B,C,D\right)\Im\Sigma_{T}^{V-V}\left(s,m_{\alpha}^{2},m_{\beta}^{2}\right)+\nonumber \\
\label{eq:11}\\
 & L_{b,\mu_{1}\mu_{2},\nu_{1}...\nu_{R}}^{1,M,1}\left(B,C,D\right)\Im\Sigma_{L}^{V-V}\left(s,m_{\alpha}^{2},m_{\beta}^{2}\right)\Bigg].\nonumber 
\end{align}
Functions $L_{a,\mu_{1}\mu_{2},\nu_{1}...\nu_{R}}$ and $L_{b,\mu_{1}\mu_{2},\nu_{1}...\nu_{R}}$
depend on two-, three- and four-points tensor coefficient functions
for the topologies in Fig.(\ref{f1}), defined from left to right,
respectively. Each of the many point functions are dependent on the
integration parameter $s$, masses and combinations of external momenta.
For the two-loop result with fermion self-energy insertions, we can
write:
\begin{align}
I_{\nu_{1}...\nu_{R}}^{1,M,1} & =\frac{1}{\pi}\stackrel[\alpha,\beta]{}{\varSigma}\stackrel[\left(m_{\alpha}+m_{\beta}\right)^{2}]{\infty}{\int}ds\cdot\Bigg[N_{a,\nu_{1}...\nu_{R}}^{1,M,1}\left(B,C,D\right)\Im\Sigma_{L}^{f}\left(s,m_{\alpha}^{2},m_{\beta}^{2}\right)\omega_{-}+\nonumber \\
\label{eq:12}\\
 & N_{b,\nu_{1}...\nu_{R}}^{1,M,1}\left(B,C,D\right)\Im\Sigma_{R}^{f}\left(s,m_{\alpha}^{2},m_{\beta}^{2}\right)\omega_{+}+N_{c,\nu_{1}...\nu_{R}}^{1,M,1}\left(B,C,D\right)\Im\Sigma_{S}^{f}\left(s,m_{\alpha}^{2},m_{\beta}^{2}\right)\Bigg].\nonumber 
\end{align}
As in the previous case, the functions $N_{a,b,c,\nu_{1}...\nu_{R}}^{1,M,1}$
are defined through functions $B,C$ and $D$ in similar fashion.
Integration in Eqs.(\ref{eq:11}) and (\ref{eq:12}) can be completed
numerically after subtraction of the UV divergences.
In case where it is possible to perform sub-loop subtraction (i.e. there
is no global UV divergence), we can use the self-energy sub-loop which has 
already subtracted terms. For example, $\gamma-\gamma$ self-energy
is: 
\begin{align}
 & \hat{\Sigma}^{\gamma-\gamma}(q^{2})=\Sigma^{\gamma-\gamma}(q^{2})-\Sigma^{\gamma-\gamma}(0)-\frac{\partial}{\partial q^{2}}\Sigma^{\gamma-\gamma}(q^{2})\Bigr|_{q^{2}=0}q^{2}=\nonumber \\
\label{eq:13}\\
 & \frac{q^{4}}{\pi}\stackrel[\alpha,\beta]{}{\varSigma}\stackrel[\left(m_{\alpha}+m_{\beta}\right)^{2}]{\infty}{\int}ds\frac{\Im\Sigma^{\gamma-\gamma}(s,m_{\alpha},m_{\beta})}{s^{2}\left(s-q^{2}-i\epsilon\right)}.\nonumber 
\end{align}
Clearly, $\Sigma^{\gamma-\gamma}(0) = 0$, but we keep this term in Eq.(\ref{eq:13}) anyway for the dispersive subtraction. This way, the second-loop integral in Eq.(\ref{eq:6}) will acquire term $\frac{q^4}{s^2}$, which will cause the cancellation of two $\gamma$ propagators around $\gamma-\gamma$ insertion, and hence we can omit use of partial fraction expansion in Eq.(\ref{eq:10a}) and still write the final second-loop integration in the form of $B,C$ and $D$ functions for the topologies defined from left to right in Fig.(\ref{f1}).
The same can be done for $Z-Z$, $\gamma-Z$ and $W-W$ mixings insertions.
For $Z-Z$ and $W-W$ insertions, we can write, in the on-shell renormalization
scheme, the following \cite{Denner}:
\begin{align}
 & \hat{\Sigma}^{V-V}(q^{2})=\Sigma^{V-V}(q^{2})-\Sigma^{V-V}(m_{V}^{2})-\frac{\partial}{\partial q^{2}}\Sigma^{V-V}(q^{2})\Bigr|_{q^{2}=m_{V}^{2}}\left(q^{2}-m_{V}^{2}\right)=\nonumber \\
\label{eq:14}\\
 & \frac{\left(q^{2}-m_{V}^{2}\right)^{2}}{\pi}\stackrel[\alpha,\beta]{}{\varSigma}\stackrel[\left(m_{\alpha}+m_{\beta}\right)^{2}]{\infty}{\int}ds\frac{\Im\Sigma^{V-V}(s,m_{\alpha},m_{\beta})}{\left(s-m_{V}^{2}\right)^{2}\left(s-q^{2}-i\epsilon\right)}.\nonumber 
\end{align}
Here, superscript $V-V$ corresponds to either $Z-Z$ or $W-W$ mixings.
For $\gamma-Z$ mixing we have:
\begin{align}
 & \hat{\Sigma}^{\gamma-Z}(q^{2})=\Sigma^{\gamma-Z}(q^{2})-\frac{1}{m_{Z}^{2}}\left[\Sigma^{\gamma-Z}(0)q^{2}-\Sigma^{\gamma-Z}(m_{Z}^{2})\left(q^{2}-m_{Z}^{2}\right)\right]=\nonumber \\
\label{eq:15}\\
 & \frac{q^{2}\left(q^{2}-m_{Z}^{2}\right)}{\pi}\stackrel[\alpha,\beta]{}{\varSigma}\stackrel[\left(m_{\alpha}+m_{\beta}\right)^{2}]{\infty}{\int}ds\frac{\Im\Sigma^{\gamma-Z}(s,m_{\alpha},m_{\beta})}{s\left(s-m_{Z}^{2}\right)\left(s-q^{2}-i\epsilon\right)}.\nonumber 
\end{align}
Eqs.(\ref{eq:14}) and (\ref{eq:15}) would cancel $Z-Z$ or $W-W$
and $\gamma-Z$ propagators in the second loop integration, and, as
a result, functions $L_{a,b}$ in Eq.(\ref{eq:11}) will again depend
on $B,C$ and $D$ Passarino-Veltman tensor coefficient functions.
In the case of $f-f$ sub-loop, we can apply the following on-shell
subtractions:
\begin{align}
 & \hat{\Sigma}^{f}(q)=\Sigma^{f}(\cancel{q})-\Sigma^{f}(m_{f})-\frac{\partial}{\partial\cancel{q}}\Sigma^{f}(\cancel{q})\Bigr|_{\cancel{q}=m_{f}}(\cancel{q}-m_{f})=\nonumber \\
\label{eq:16}\\
 & \cancel{q}\omega_{-}\left(I_{L}+a_{L}\right)+\cancel{q}\omega_{+}\left(I_{R}+a_{R}\right)+m_{f}\left(I_{S}+a_{S}\right).\nonumber 
\end{align}
Here, the functions $I_{L,R,S}$ have the following integral representation:
\begin{align}
 & I_{L,R,S}=\frac{q^{2}-m_{f}^{2}}{\pi}\stackrel[\alpha,\beta]{}{\varSigma}\stackrel[\left(m_{\alpha}+m_{\beta}\right)^{2}]{\infty}{\int}ds\frac{\Im\Sigma_{L,R,S}^{f}(s,m_{\alpha},m_{\beta})}{\left(s-m_{f}^{2}\right)\left(s-q^{2}-i\epsilon\right)}\nonumber \\
 & \textrm{and}\label{eq:16-a}\\
 & a_{L,R}=-2m_{f}^{2}\left(\Sigma'_{L,R}\left(m_{f}^{2}\right)+\Sigma'_{S}\left(m_{f}^{2}\right)\right)\nonumber \\
\nonumber \\
 & a_{S}=m_{f}^{2}\left(\Sigma'_{L}\left(m_{f}^{2}\right)+\Sigma'_{R}\left(m_{f}^{2}\right)+2\Sigma'_{S}\left(m_{f}^{2}\right)\right).\nonumber 
\end{align}
Substitution of Eq.(\ref{eq:16}) into the second-loop integration will
result in the cancelation of $\left(q^{2}-m_{f}^{2}\right)$ in $I_{L,R,S}$
with one of the fermion propagators. Terms $a_{L,R,S}$ do not result
in cancellations, but they also do not introduce the dispersion denominator
$\left(s-q^{2}-i\epsilon\right)^{-1}$ into the second-loop integration. After applying partial fraction expansion of the denominators with the same momentum, functions $N_{a,b,c}$ would depend on $B,C$ and $D$ tensor coefficient functions only.

The structure of the insertions in Eq.(\ref{eq:6}) suggests that we can
introduce $V-V$ and $f-f$ effective mixing propagators with a dispersion
integral removed. All the functions of parameter $s$ could be left
un-evaluated during the second-loop integration. This gives us the possibility
to employ a computer-algebra approach, where the second-loop integral
could be evaluated analytically, and after subtractions the dispersion
integration can be carried out numerically. For $V-V$ effective mixing
we can write that as a combination of the transverse and longitudinal
propagators:
\begin{align}
 & \Pi_{\mu\nu}^{V-V}(q)=\Pi_{T,\mu\nu}^{V-V}+\Pi_{L,\mu\nu}^{V-V}\label{eq:17}\\
\nonumber \\
 & \Pi_{T,\mu\nu}^{V-V}=\frac{-ig_{\rho\mu}}{q^{2}-m_{V}^{2}}\left[\frac{g^{\rho\sigma}-\frac{q^{\rho}q^{\sigma}}{q^{2}}}{s-q^{2}-i\epsilon}\Im\Sigma_{T}^{V-V}\left(s,m_{\alpha}^{2},m_{\beta}^{2}\right)\right]\frac{-ig_{\sigma\nu}}{q^{2}-m_{V}^{2}}\nonumber \\
\nonumber \\
 & \Pi_{L,\mu\nu}^{V-V}=\frac{-ig_{\rho\mu}}{q^{2}-m_{V}^{2}}\left[\frac{\frac{q^{\rho}q^{\sigma}}{q^{2}}}{s-q^{2}-i\epsilon}\Im\Sigma_{L}^{V-V}\left(s,m_{\alpha}^{2},m_{\beta}^{2}\right)\right]\frac{-ig_{\sigma\nu}}{q^{2}-m_{V}^{2}}.\nonumber 
\end{align}
When evaluating the second-loop integral we can leave imaginary parts
of $\Sigma_{T,L}$ un-evaluated and get analytical structure for the
two-loop graph. In case if subtraction is possible at the sub-loop
level, $V-V$ effective propagators would have the following structure:
\begin{align}
 & \hat{\Pi}_{\mu\nu}^{V-V}(q)=\hat{\Pi}_{T,\mu\nu}^{V-V}+\hat{\Pi}_{L,\mu\nu}^{V-V}\label{eq:18}\\
\nonumber \\
 & \hat{\Pi}_{T,\mu\nu}^{V-V}=-T^{V-V}\left(s,m_{V}^{2}\right)\left[\frac{g_{\mu\nu}-\frac{q_{\mu}q_{\nu}}{q^{2}}}{s-q^{2}-i\epsilon}\right]\Im\Sigma_{T}^{V-V}\left(s,m_{\alpha}^{2},m_{\beta}^{2}\right)\nonumber \\
\nonumber \\
 & \hat{\Pi}_{L,\mu\nu}^{V-V}=-T^{V-V}\left(s,m_{V}^{2}\right)\left[\frac{\frac{q_{\mu}q_{\nu}}{q^{2}}}{s-q^{2}-i\epsilon}\right]\Im\Sigma_{L}^{V-V}\left(s,m_{\alpha}^{2},m_{\beta}^{2}\right).\nonumber 
\end{align}
 Here, the functions $T^{V-V}\left(s,m_{V}^{2}\right)$ and $\Im\Sigma_{T,L}$
are independent of the second-loop momenta and could be left un-evaluated
until dispersion integration is performed. For the specific $V-V$
mixings, functions in Eq.(\ref{eq:18}) are defined in Tbl.(\ref{tbl1}).
\begin{table}
\begin{centering}
\begin{tabular}{|c|c|c|c|}
\hline 
 & $\gamma-\gamma$ & $\{Z,W\}-\{Z,W\}$ & $\gamma-Z$\tabularnewline
\hline 
$T^{V-V}$ & ${\displaystyle \frac{1}{s^{2}}}$ & ${\displaystyle \frac{1}{\left(s-m_{\{Z,W\}}^{2}\right)^{2}}}$ & ${\displaystyle \frac{1}{s\left(s-m_{z}^{2}\right)}}$\tabularnewline
\hline 
\end{tabular}
\par\end{centering}
\caption{Structures of the function $T^{V-V}\left(q^{2},s,m_{V}^{2}\right)$
for specific $V-V$ mixings.}

\label{tbl1}
\end{table}
 For fermion mixing, we can also introduce following effective propagator
\begin{align}
 & \Pi^{f}(q)=\frac{1}{\cancel{q}-m_{f}}\left[\frac{G\left(q,s,m_{\alpha},m_{\beta}\right)}{s-q^{2}-i\epsilon}\right]\frac{1}{\cancel{q}-m_{f}}.\label{eq:19}
\end{align}
For the subtracted $f-f$ sub-loop insertions, we can replace Eq.(\ref{eq:19})
and introduce, with the help of Eqs.(\ref{eq:16}) and (\ref{eq:16-a}),
the following set of effective fermion propagators derived from the
first-loop integration: 
\begin{align}
 & \hat{\Pi}^{f}(q)=\hat{\Pi}_{1}^{f}(q)+\hat{\Pi}_{2}^{f}(q)\label{eq:20}\\
\nonumber \\
 & \hat{\Pi}_{1}^{f}(q)=\left(\cancel{q}+m_{f}\right)\left[\frac{y_{L}\cancel{q}\omega_{-}+y_{R}\cancel{q}\omega_{+}+m_{f}y_{S}}{\left(q^{2}-m_{f}^{2}\right)\left(s-q^{2}-i\epsilon\right)}\right]\left(\cancel{q}+m_{f}\right)\nonumber \\
\nonumber \\
 & \hat{\Pi}_{2}^{f}(q)=\frac{1}{\cancel{q}-m_{f}}\left[d_{L}\cancel{q}\omega_{-}+d_{R}\cancel{q}\omega_{+}+m_{f}d_{S}\right]\frac{1}{\cancel{q}-m_{f}},\nonumber 
\end{align}
where 
\begin{align*}
 & y_{L,R,S}\equiv y_{L,R,S}\left(s,m_{\alpha}^{2},m_{\beta}^{2}\right)=\frac{\Im\Sigma_{L,R,S}^{f}}{s-m_{f}^{2}}.
\end{align*}
The UV-finite functions $d_{L,R,S}\left(s,m_{\alpha}^{2},m_{\beta}^{2}\right)$
can be derived from Eq.(\ref{eq:16-a}) using a dispersive representation
of the constants $a_{L,R,S}$:
\begin{align*}
 & d_{L,R}\equiv d_{L,R}\left(s,m_{\alpha}^{2},m_{\beta}^{2}\right)=-2m_{f}^{2}\frac{\Im\Sigma_{L,R}^{f}+\Im\Sigma_{S}^{f}}{\left(s-m_{f}^{2}\right)^{2}}\\
\\
 & d_{S}\equiv d_{S}\left(s,m_{\alpha}^{2},m_{\beta}^{2}\right)=m_{f}^{2}\frac{\Im\Sigma_{L}^{f}+\Im\Sigma_{R}^{f}+2\Im\Sigma_{S}^{f}}{\left(s-m_{f}^{2}\right)^{2}},
\end{align*}
where $\Im\Sigma_{L,R,S}^{f}\equiv\Im\Sigma_{L,R,S}^{f}\left(s,m_{\alpha}^{2},m_{\beta}^{2}\right)$.
The implementation of the effective propagators in Eqs.(\ref{eq:17},
\ref{eq:18}, \ref{eq:19}) and (\ref{eq:20}) is straightforward
in the computer-algebra packages such as FormCalc or Form (\cite{LoopTools}
and \cite{FORM}), and as a result it is possible to construct the
two-loop self-energies matrix elements in the analytical form. If
additional subtractions are needed, it could be done later using the
second-order EW counterterms (\cite{Hollik-2} and \cite{Freitas}).
And, finally, the last step of the calculations would be numerical
integration. This can be done with the help of the numerical libraries
from LoopTools or FF and integration routines such as VEGAS or QUADPACK. 
Since the dispersion integration would involve many-point tensor coefficient
functions (above the two-point functions), a numerical stability could
become a concern. Out of all the Passarino-Veltman functions, only
two-point tensor coefficient functions have well-defined analytical
structure and therefore the most stable numerically. It would be most
desirable if we could write three-, four- and five-point functions (for the cases of triangle type of the sub-loop insertion)
which enter functions $L_{a,b}$ and $N_{a,b,c}$ in Eqs.(\ref{eq:11})
and (\ref{eq:12}) as some representation of the two-point tensor
coefficient functions. This can be achieved if we combine Feynman
trick \cite{Freitas,Ghinculov} with derivative representation of
the many-point functions. Let us now consider three-, four- and five-point
functions separately. 

We start with the scalar three-point function and later consider results
for the $C_{i,ij,ijk}$ Passarino-Veltman functions. The general expression
for the three-point scalar function is given by:
\begin{align}
 & C_{0}\equiv C_{0}\left(p_{1}^{2},p_{2}^{2},\left(p_{1}+p_{2}\right)^{2},m_{1}^{2},m_{2}^{2},m_{3}^{2}\right)=\nonumber \\
\label{eq:21}\\
 & \frac{\mu^{4-D}}{i\pi^{D/2}}\int d^{D}q\frac{1}{\left[q^{2}-m_{1}^{2}\right]\left[\left(q+p_{1}\right)^{2}-m_{2}^{2}\right]\left[\left(q+p_{1}+p_{2}\right)^{2}-m_{3}^{2}\right]}.\nonumber 
\end{align}
With the help of Feynman trick, we can join the first two propagators
in Eq.\ref{eq:21}, and after shifting momentum $q=\tau-p_{1}-p_{2}$,
we can write
\begin{align}
 & C_{0}=\frac{\mu^{4-D}}{i\pi^{D/2}}\intop_{0}^{1}dx\int d^{D}\tau\frac{1}{\left[\left(\tau-\left(p_{1}\bar{x}+p_{2}\right)\right)^{2}-m_{12}^{2}\right]^{2}\left[\tau^{2}-m_{3}^{2}\right]}\nonumber \\
\label{eq:22}\\
 & m_{12}^{2}=m_{1}^{2}\bar{x}+m_{2}^{2}x-p_{1}^{2}x\bar{x},\nonumber 
\end{align}
where $\bar{x}=1-x$. Term $\left[\left(\tau-\left(p_{1}\bar{x}+p_{2}\right)\right)^{2}-m_{12}^{2}\right]^{-2}$
can be replaced after shifting mass $m_{12}^{2}$ by a small parameter
$\lambda$.
\begin{align}
 & \frac{1}{\left[\left(\tau-\left(p_{1}\bar{x}+p_{2}\right)\right)^{2}-m_{12}^{2}\right]^2}=\underset{\lambda\rightarrow0}{\lim}\frac{\partial}{\partial\lambda}\frac{1}{\left[\left(\tau-\left(p_{1}\bar{x}+p_{2}\right)\right)^{2}-\left(m_{12}^{2}+\lambda\right)\right]}
\label{eq:22a}
\end{align}
As a result, the expression for the three-point function can be reduced
to the derivative representation of two-point function (see Fig.(\ref{f3})):
\begin{figure}
\begin{centering}
\includegraphics[scale=0.55]{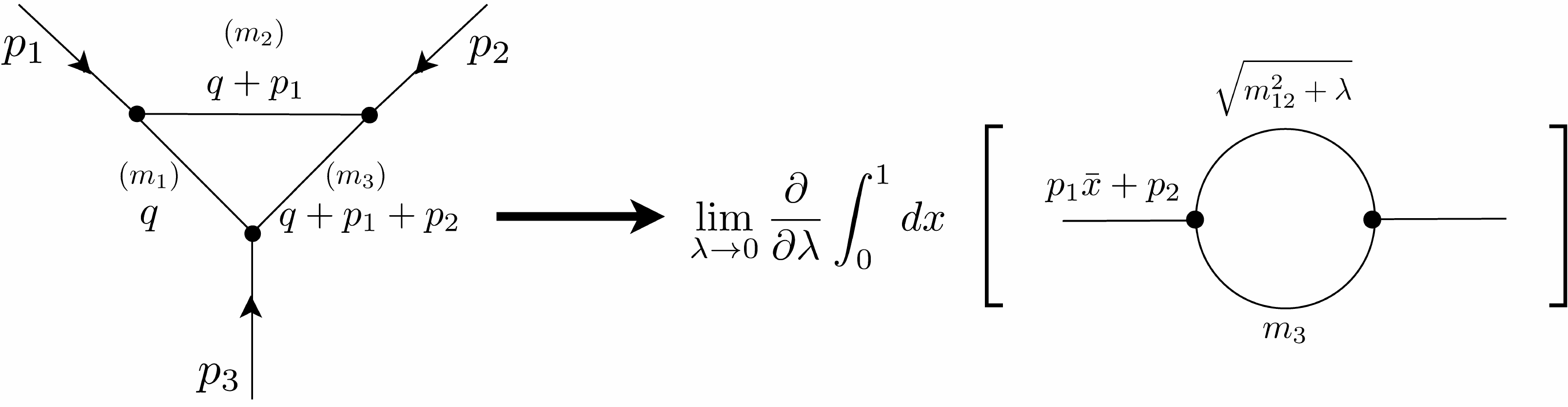}
\par\end{centering}
\caption{Reduction of the triangle graph by the derivative representation of
self-energy.}

\label{f3}
\end{figure}
\begin{align}
 & C_{0}=&\frac{\mu^{4-D}}{i\pi^{D/2}}\lim_{\lambda\rightarrow0}\frac{\partial}{\partial\lambda}\intop_{0}^{1}dx\int d^{D}\tau\frac{1}{\left[\left(\tau-\left(p_{1}\bar{x}+p_{2}\right)\right)^{2}-\left(m_{12}^{2}+\lambda\right)\right]\left[\tau^{2}-m_{3}^{2}\right]}=\nonumber \\
\nonumber \\
 & &\lim_{\lambda\rightarrow0}\frac{\partial}{\partial\lambda}\intop_{0}^{1}dx\;B_{0}\left(\left(p_{1}\bar{x}+p_{2}\right)^{2},m_{3}^{2},m_{12}^{2}+\lambda\right).\label{eq:23}
\end{align}
On the contrary, in \cite{Ghinculov} the differentiation in Eq.(\ref{eq:23})
would be done with respect to $m_{12}^{2}$, but this would require
the analytical differentiation first (since $m_{12}^{2}$ is a function
of the Feynman parameter), and then integration over the Feynman parameter.
In our case, the integration over the Feynman parameter, and then
differentiation with respect to $\lambda$ can be all done numerically.
Using the approach from Eq.(\ref{eq:23}), we can replace the many-point
tensor functions in the second-loop integration by the derivatives
of two-point function. Taking into account that the LoopTools
libraries have numerical implementation for the regularization of UV and
IR divergences, we can perform the two-loop calculations using the
two-point functions basis and later, after the appropriate subtractions,
perform the dispersive and the Feynman integration numerically. The
final step would be a numerical evaluation of the derivative of the integrated
result. Since LoopTools and FF libraries are capable of computing
tensor coefficient functions of the higher-rank tensors, we can derive
the partial tensor reduction for the three-point function in terms
of the two-point basis. For the function $C_{\mu_{1}...\mu_{N}}$,
we can write:
\begin{align}
 & C_{\mu_{1}...\mu_{N}}=\frac{\mu^{4-D}}{i\pi^{D/2}}\int d^{D}q\frac{q_{\mu_{1}}...q_{\mu_{N}}}{\left[q^{2}-m_{1}^{2}\right]\left[\left(q+p_{1}\right)^{2}-m_{2}^{2}\right]\left[\left(q+p_{1}+p_{2}\right)^{2}-m_{3}^{2}\right]}.\label{eq:24}
\end{align}
 For the vector $C_{\mu}$, we can use the tensor decomposition and
in parallel apply the Feynman trick on the right hand side of Eq.(\ref{eq:24}).
After using the derivative approach from Eq.(\ref{eq:23}), and shifting
the momenta as before, we get:
\begin{align}
 & p_{1\mu}C_{1}+\left(p_{1\mu}+p_{2\mu}\right)C_{2}=\lim_{\lambda\rightarrow0}\frac{\partial}{\partial\lambda}\intop_{0}^{1}dx\left[B_{\mu}-\left(p_{1\mu}+p_{2\mu}\right)B_{0}\right],\label{eq:25}
\end{align}
where $B_{0,\mu}\equiv B_{0,\mu}\left(\left(p_{1}\bar{x}+p_{2}\right)^{2},m_{3}^{2},m_{12}^{2}+\lambda\right)$.
Using the tensor decomposition for $B_{\mu}=-\left(p_{1\mu}\bar{x}+p_{2\mu}\right)B_{1}$,
and matching terms in front of $p_{1\mu}$ and $p_{2\mu}$, we can
solve for $C_{1,2}$ in terms of $B_{0,1}$ functions:
\begin{align}
 & C_{1}=\lim_{\lambda\rightarrow0}\frac{\partial}{\partial\lambda}\intop_{0}^{1}dxB_{1}x\nonumber \\
\label{eq:26}\\
 & C_{2}=-\lim_{\lambda\rightarrow0}\frac{\partial}{\partial\lambda}\intop_{0}^{1}dx\left[B_{0}+B_{1}\right].\nonumber 
\end{align}
The same idea can be extrapolated to the higher orders of the three-point
functions. The results for the $C_{\mu\nu}$ and $C_{\mu\nu\alpha}$
partial reduction are given in Appendix A. An important advantage
of the partial tensor reduction is that we can substantially reduce
the size of the final expressions in the two-loop integrals. The reduction
of the three-point functions to the two-point basis can be also employed
in the dispersive representation of $C_{0,\mu,\mu\nu,\mu\nu\alpha}$ which
could prove helpful if we use the triangle sub-loop insertion. 

For the four-point function, we will use an analogous approach. We
start with the general structure of scalar $D_{0}$ function:
\begin{align*}
 & D_{0}\equiv D_{0}\left(p_{1}^{2},p_{2}^{2},p_{3}^{2},p_{4}^{2},\left(p_{1}+p_{2}\right)^{2},\left(p_{2}+p_{3}\right)^{2},m_{1}^{2},m_{2}^{2},m_{3}^{2},m_{4}^{2}\right)=\\
\\
 & \frac{\mu^{4-D}}{i\pi^{D/2}}\int d^{D}q\frac{1}{\left[q^{2}-m_{1}^{2}\right]\left[\left(q+p_{1}\right)^{2}-m_{2}^{2}\right]\left[\left(q+p_{1}+p_{2}\right)^{2}-m_{3}^{2}\right]\left[\left(q+p_{1}+p_{2}+p_{3}\right)^{2}-m_{4}^{2}\right]}.
\end{align*}
In this case, we join the first three propagators, and after shifting
momentum $q=\tau-\stackrel[i=1]{3}{\Sigma}p_{i}$, we get:
\begin{align}
 & D_{0}=2\frac{\mu^{4-D}}{i\pi^{D/2}}\intop_{0}^{1}dx\intop_{0}^{1-x}dy\int d^{D}\tau\frac{1}{\left[\left(\tau-\left(p_{1}\left(\bar{x}-y\right)+p_{2}\bar{y}+p_{3}\right)\right)^{2}-m_{123}^{2}\right]^{3}\left[\tau^{2}-m_{4}^{2}\right]}\nonumber \\
\label{eq:27}\\
 & m_{123}^{2}=m{}_{1}^{2}\left(\bar{x}-y\right)+m_{2}^{2}x+m_{3}^{2}y-p_{1}^{2}x\bar{x}-p_{12}^{2}y\bar{y}+2xy\left(p_{1}p_{12}\right)\nonumber \\
\nonumber \\
 & p_{12}=p_{1}+p_{2}.\nonumber 
\end{align}
Obviously, the reduction to the two-point $B_{0}$ function is achieved
by the second-order differentiation with respect to mass shift parameter
$\lambda$:
\begin{align}
 & D_{0}=\lim_{\lambda\rightarrow0}\frac{\partial^{2}}{\partial\lambda^{2}}\intop_{0}^{1}dx\intop_{0}^{1-x}dy\, B_{0}\left[\left(p_{1}\left(\bar{x}-y\right)+p_{2}\bar{y}+p_{3}\right)^{2},m_{4}^{2},m_{123}^{2}+\lambda\right].\label{eq:28}
\end{align}
Again, the partial reduction for $D_{\mu,\mu\nu,\mu\nu\rho,\mu\nu\rho\sigma}$
can be done in the similar way as in Eq.(\ref{eq:25}) (Appendix B).

The five-point functions are reduced as before with the help of the Feynman
trick, the third-order differentiation with respect to the mass shift
parameter $\lambda$, and shift of momentum $q=\tau-\Sigma_{i=1}^{4}p_{i}$.
Here, we can write
\begin{align}
& E_{0}\equiv E_{0}\left(p_{1}^{2},p_{2}^{2},p_{3}^{2},p_{4}^{2},p_{5}^{2},p_{12}^{2},p_{23}^{2},p_{34}^{2},p_{45}^{2},p_{51}^{2},m_{1}^{2},m_{2}^{2},m_{3}^{2},m_{4}^{2},m_{5}^{2}\right)&=\nonumber\\
\nonumber\\
& \frac{\mu^{4-D}}{i\pi^{D/2}}\intop_{0}^{1}dx\intop_{0}^{1-x}dy\intop_{0}^{1-x-y}dz\int \frac{d^{D}\tau}{\left[\left(\tau-\left(p_{1}\left(\bar{x}-y-z\right)+p_{2}\left(\bar{y}-z\right)+p_{3}\bar{z}+p_{4}\right)\right)^{2}-\left(m_{1234}^{2}+\lambda\right)\right]^{4}}\cdot\nonumber \\ 
\nonumber \\ 
& \frac{1}{\left[\tau^{2}-m_{5}^{2}\right]}=
\lim_{\lambda\rightarrow0}\frac{\partial^{3}}{\partial\lambda^{3}}\intop_{0}^{1}dx\intop_{0}^{1-x}dy\intop_{0}^{1-x-y}dz\,B_{0}\left[\left(p_{1}\left(\bar{x}-y-z\right)+p_{2}\left(\bar{y}-z\right)+p_{3}\bar{z}+p_{4}\right)^{2},m_{5}^{2},m_{1234}^{2}+\lambda\right],\label{eq:28a}
\end{align}
where $m_{1234}^{2}=m_{1}^{2}\left(\bar{x}-y-z\right)+m_{2}^{2}x+m_{3}^{2}y+m_{4}^{2}z-p_{1}^{2}\bar{x}x-p_{12}^{2}\bar{y}y-p_{123}^{2}\bar{z}z+2xy\left(p_{1}p_{12}\right)+2xz\left(p_{1}p_{123}\right)+2yz\left(p_{12}p_{123}\right)$ and $p_{ij}=p_{i}+p_{j}$, $p_{ijk}=p_{i}+p_{j}+p_{k}$.
Results of the reduction for $E$ functions are given in Appendix
C for the tensor coefficients functions up to the fourth rank. As
one can see, we can construct the two-loop integrals with the self-energy
sub-loop insertions using only the two-point basis. Employing the
effective propagators, we can now construct model files for the FeynArts
\cite{FeynArts} package and obtain the two-loop results in FormCalc
or FORM. The final UV- and IR-regularized expressions would have to
be renormalized by means of subtractions. If sub-loop subtraction
is not possible, we can regularize the insertion by a cutoff of the
dispersion integral. In this case, the renormalization constants from
counterterms responsible for the cancellations of sub-loop UV divergences,
would have to be calculated using the dispersive representation with
the same cutoff. In the last stages, numerical dispersion integration
and differentiation would have to be done. The same ideas can be extrapolated
in the case of the triangle-type of insertions in the self-energy,
vertex or box diagrams.

\subsection{Triangle Sub-Loop}

\begin{figure}
\begin{centering}
\includegraphics[bb=8cm 0bp 50cm 747bp,scale=0.35]{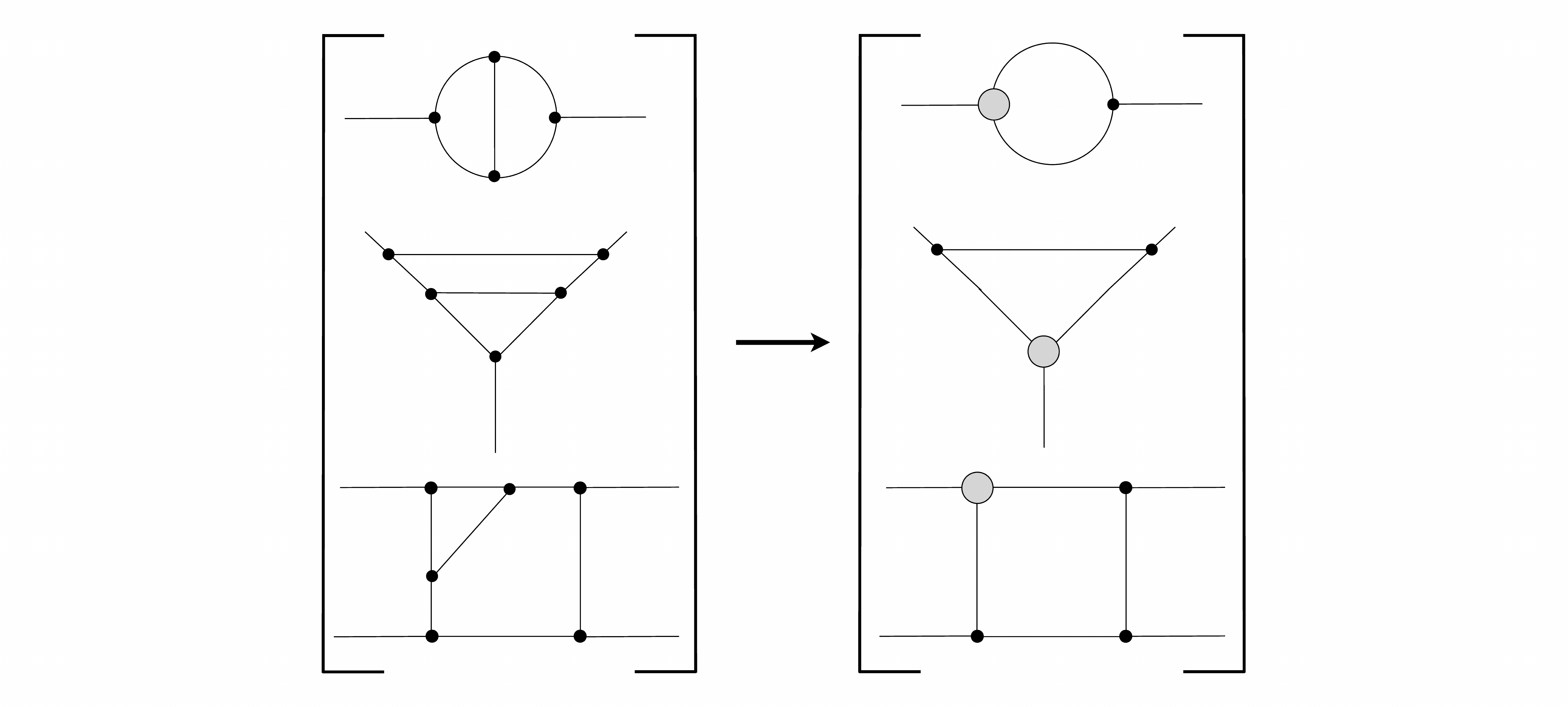}
\par\end{centering}
\caption{Examples of the triangle sub-loop in two-loops topology. In general,
triangle could be constructed around any vertex of the second loop.}

\label{fig4}
\end{figure}

\begin{figure}
\centering{}\includegraphics[scale=0.3]{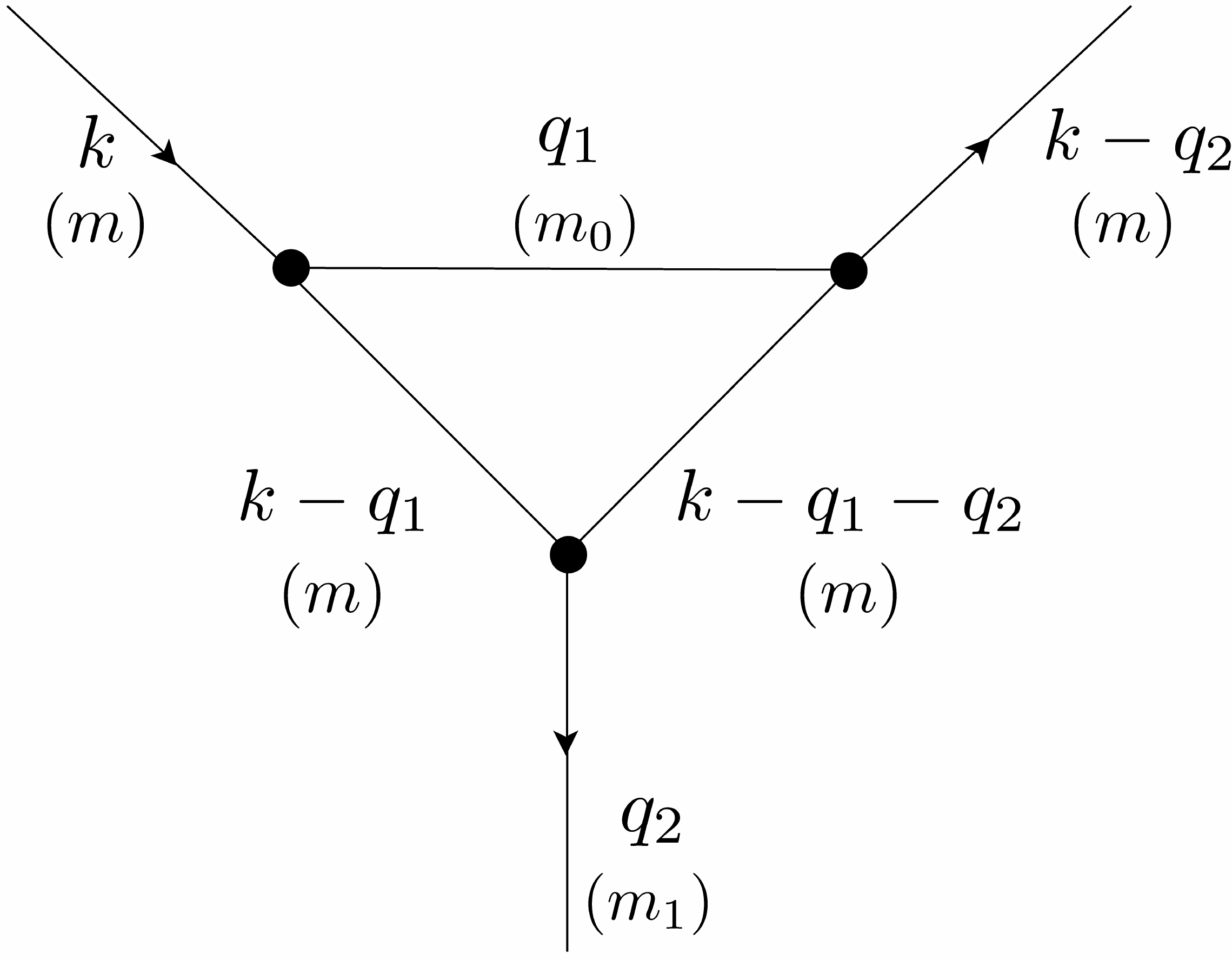}\caption{Triangle insertion with one of the legs on-shell. Here, $q_{2}$ is
the second loop integration momentum.}
\label{fig5}
\end{figure}

Examples of the triangle sub-loop insertion in two-loops topology
are shown on Fig.(\ref{fig4}). Our starting point here would be to
construct the dispersive representation of the three-point function,
which later could be used in the second-loop integration. 
To simplify, we will consider the case in which one of the external legs
of the triangle insertion is put on-shell (see Fig.(\ref{fig5})).
This could be a case shown on Fig.(\ref{fig4}), for the triangle
insertion in the box acting as the second loop. 
Considering that all particles in the loop are scalars, the graph on Fig.(\ref{fig5})
is a three-point scalar function, and using Eq.(\ref{eq:21}) notation,
we can write: 
\begin{align}
 & C_{0}\left(m^{2},q_{2}^{2},\left(q_{2}-k\right)^{2},m_{0}^{2},m^{2},m^{2}\right)=\frac{\mu^{4-D}}{i\pi^{D/2}}\int\frac{d^{D}q_{1}}{\left[q_{1}^{2}-m_{0}^{2}\right]\left[\left(q_{1}-k\right)^{2}-m^{2}\right]\left[\left(q_{1}-k+q_{2}\right)^{2}-m^{2}\right]},\label{eq:30}
\end{align}
where we used $p_{1}=-k$, $p_{2}=q_{2}$ and $p_{3}=k-q_{2}$. In
order to replace the three-point function in Eq.(\ref{eq:30}) by
the dispersive representation \cite{S-matrix}, we can again use
the Feynman trick and join the first two propagators. It is important
to apply the Feynman trick to the propagators without the second-loop 
momenta ($q_2$), otherwise it would become a part of the effective mass $m_{12}$.
If necessary, an appropriate shift of the momentum can be done to
isolate the propagators with momentum of the first loop only. Using
Eq.(\ref{eq:23}), we can write:

\begin{align}
 & C_{0}\left(m^{2},q_{2}^{2},\left(q_{2}-k\right)^{2},m_{0}^{2},m^{2},m^{2}\right)=\lim_{\lambda\rightarrow0}\frac{\partial}{\partial\lambda}\intop_{0}^{1}dx\;B_{0}\left(\left(q_{2}-k\bar{x}\right)^{2},m_{3}^{2},m_{12}^{2}+\lambda\right)\nonumber \\
\label{eq:31}\\
 & m_{12}^{2}=m_{0}^{2}\bar{x}+m^{2}x^{2}.\nonumber 
\end{align}
The two-point function can be easily written as a dispersion integral
and substituted into Eq.(\ref{eq:31}):
\begin{align}
 & C_{0}\left(m^{2},q_{2}^{2},\left(q_{2}-k\right)^{2},m_{0}^{2},m^{2},m^{2}\right)=\frac{1}{\pi}\lim_{\lambda\rightarrow0}\frac{\partial}{\partial\lambda}\intop_{0}^{1}dx\intop_{\left(m_{3}+\left(m_{12}^{2}+\lambda\right)^{1/2}\right)^{2}}^{\Lambda^{2}}ds\;\frac{\Im B_{0}\left(s,m_{3}^{2},m_{12}^{2}+\lambda\right)}{s-\left(q_{2}-k\bar{x}\right)^{2}-i\epsilon}.\label{eq:32}
\end{align}

\begin{figure}
\begin{centering}
\includegraphics[scale=0.5]{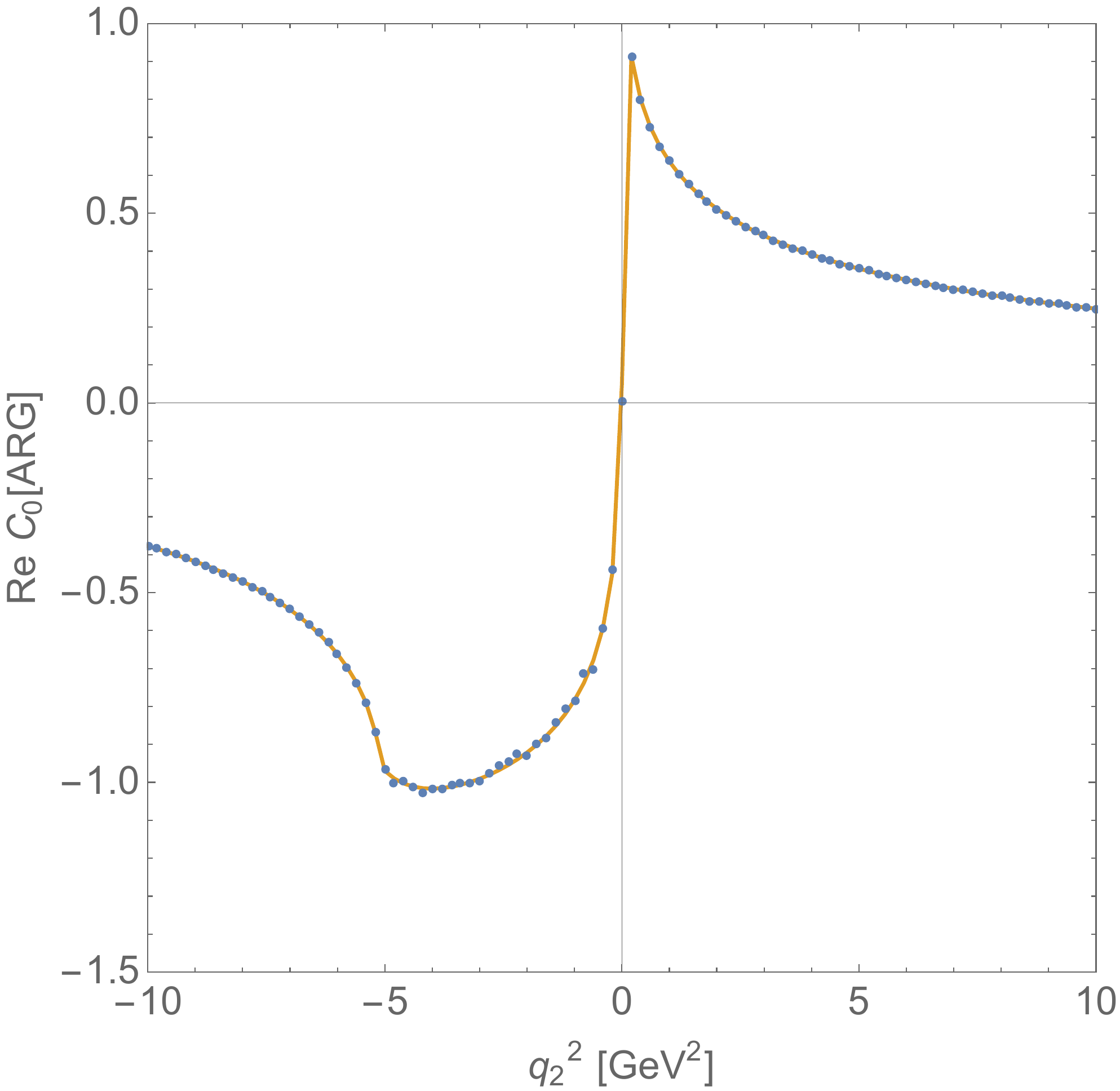}
\par\end{centering}
\caption{Comparison of the results obtained using LoopTools (solid line) and
dispersion integral in Eq.(\ref{eq:32}) (dots). Here, we have used
for $m_{0}=1.2$ GeV, $m=0.1$ GeV and $\left(k\cdot q_{2}\right)=-3.4$
GeV$^{2}$.}

\label{fig6}
\end{figure}

\begin{figure}
\begin{centering}
\includegraphics[scale=0.4]{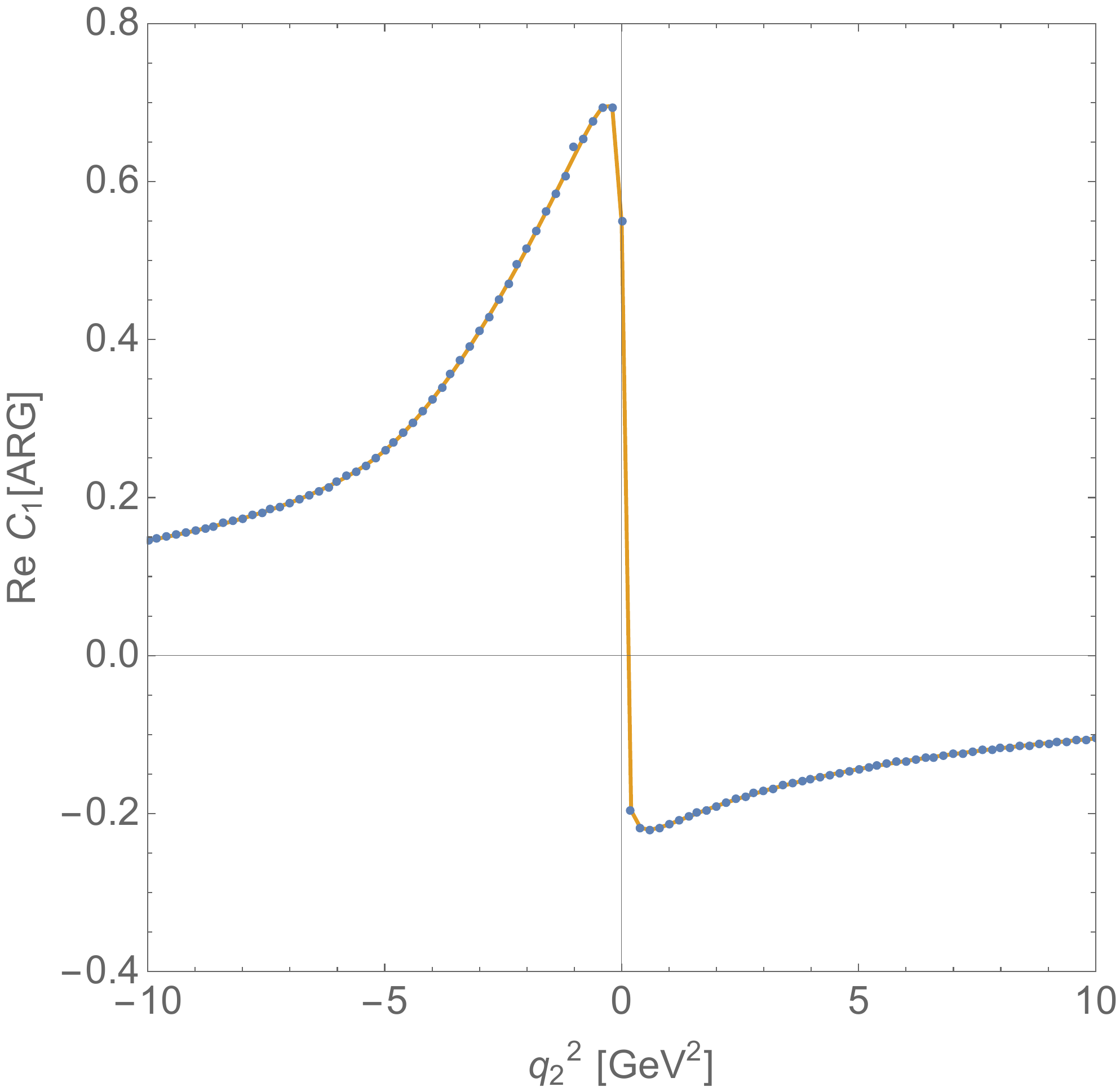} \includegraphics[scale=0.4]{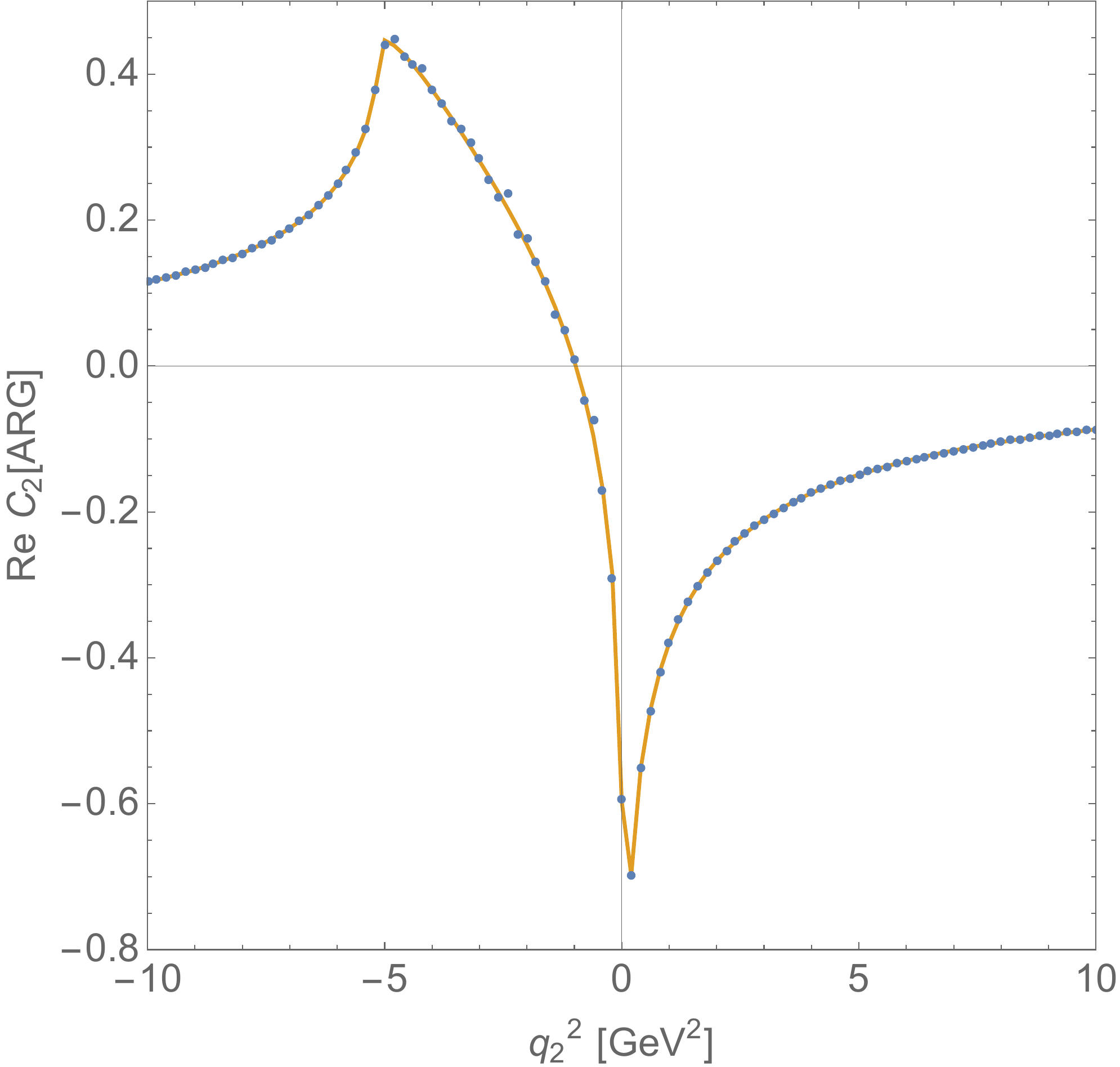}
\par\end{centering}
\caption{Comparison of the $C_{1}$ and $C_{2}$ functions calculated from
dispersion integrals (dots) and LoopTools (solid line). Masses and
external momenta have the same values as for Fig.(\ref{fig6})}

\label{fig7}
\end{figure}

\begin{figure}
\begin{centering}
\includegraphics[scale=0.4]{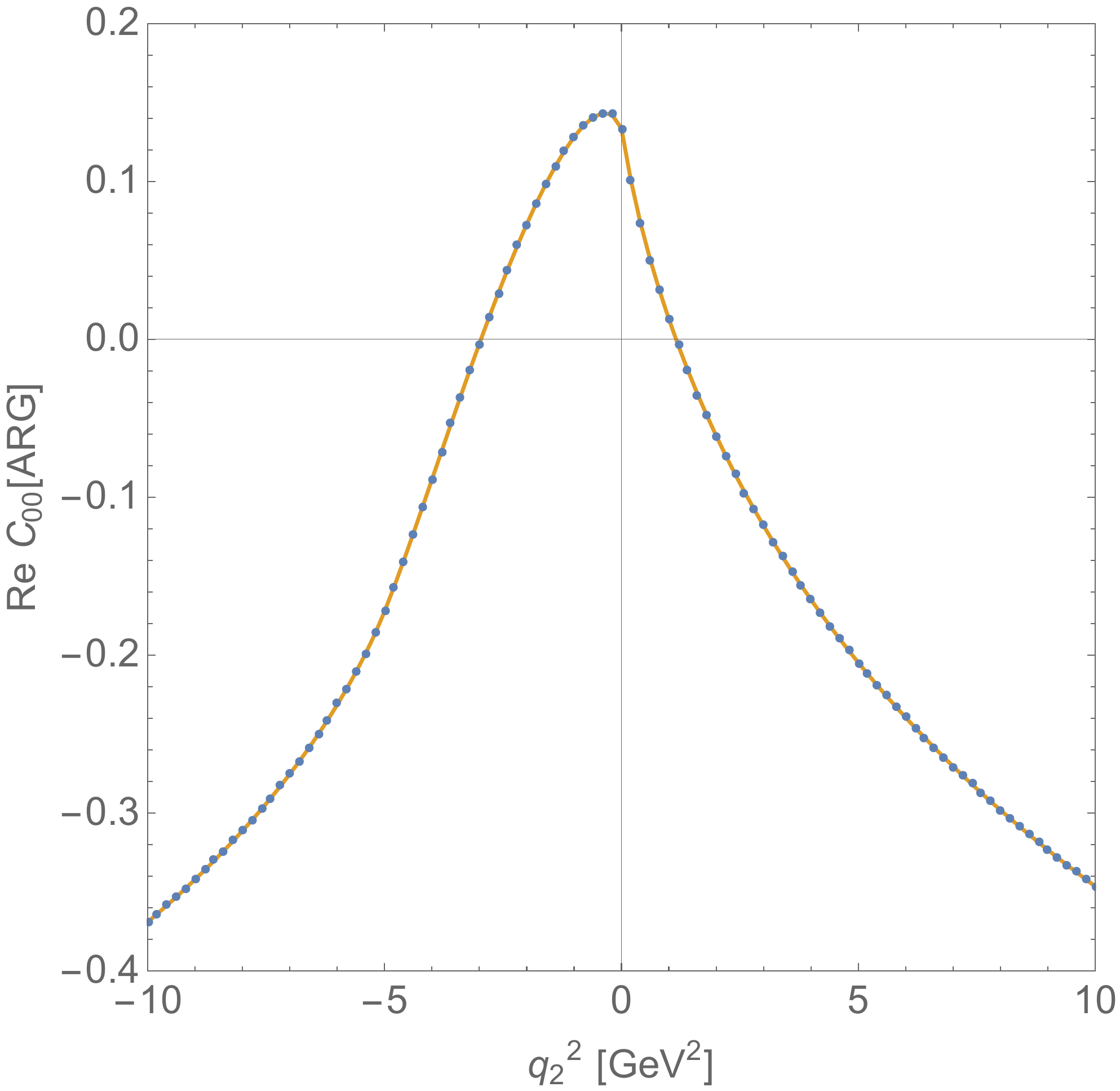} \includegraphics[scale=0.4]{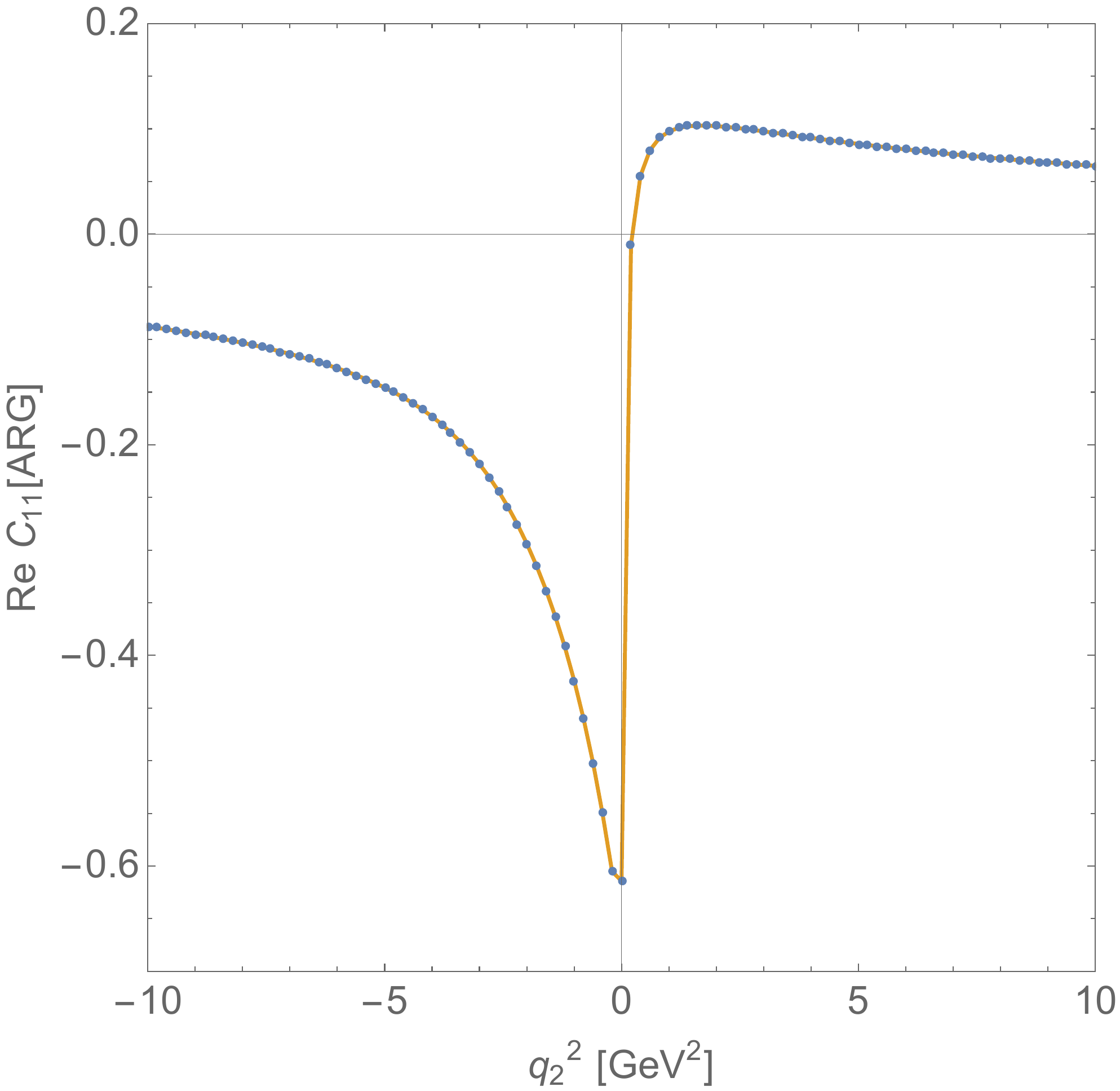}
\includegraphics[scale=0.4]{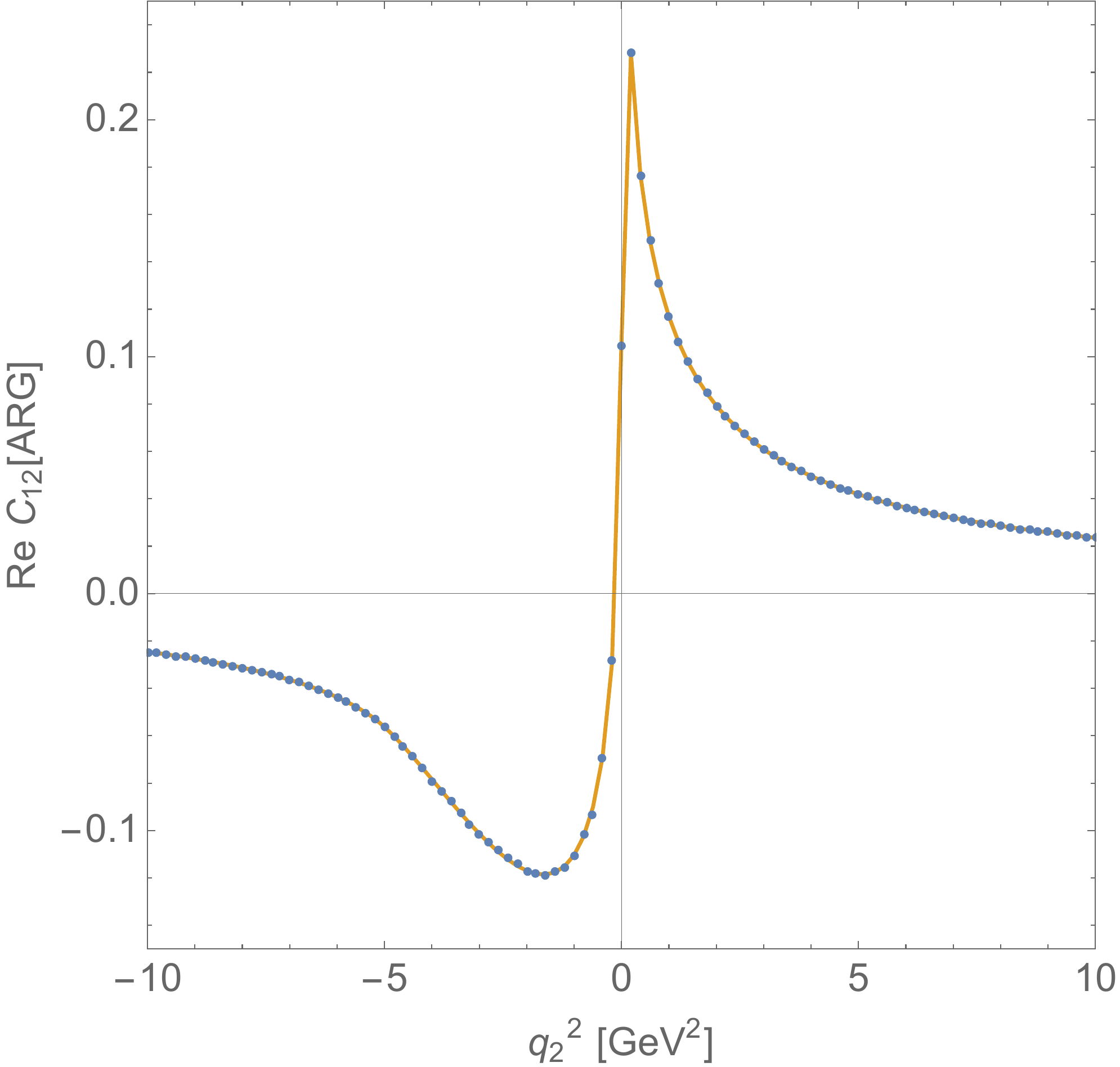} \includegraphics[scale=0.4]{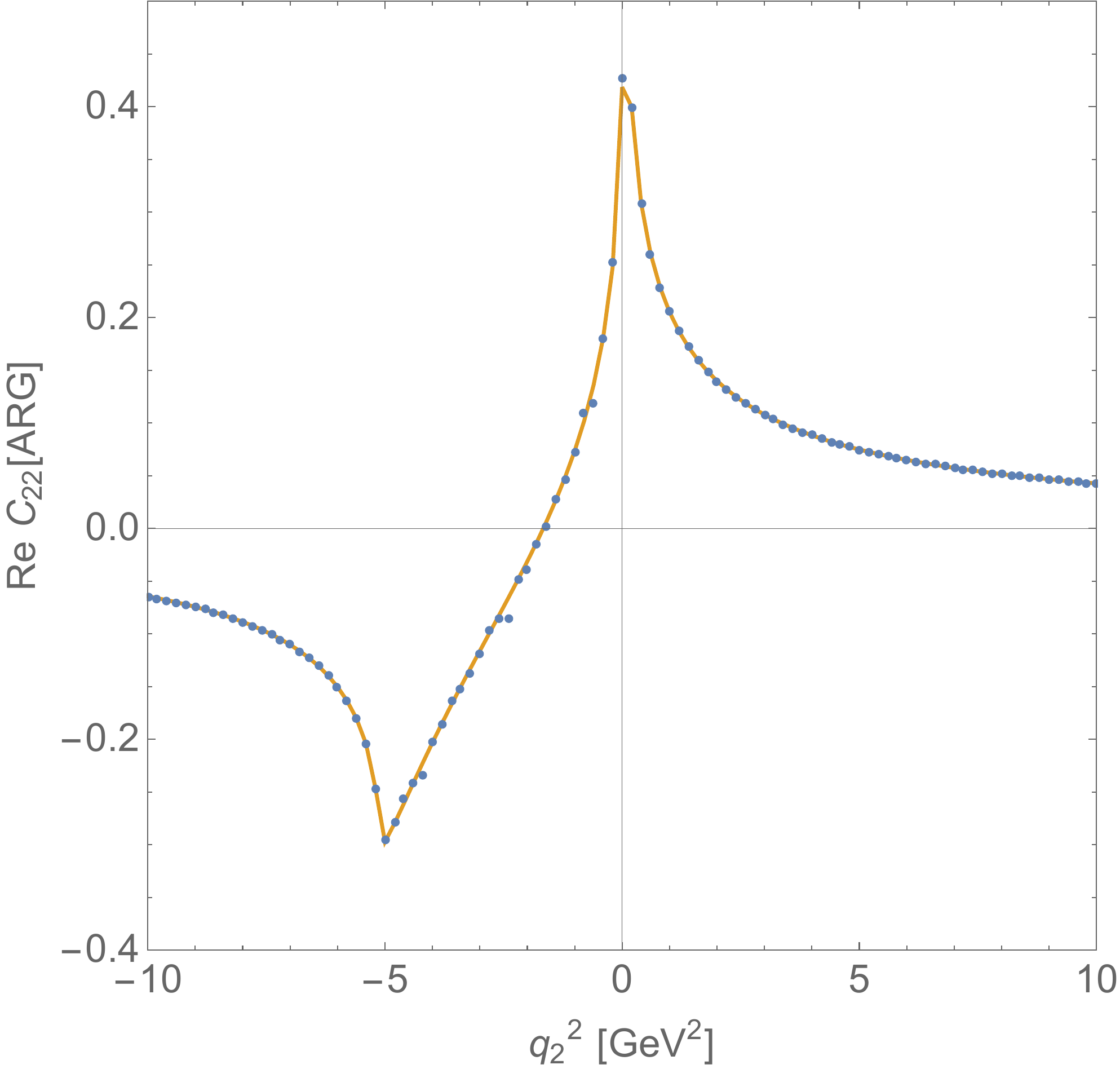}
\par\end{centering}
\caption{Comparison of the $C_{00,11,12,22}$ functions calculated from dispersion
integrals (dots) and LoopTools (solid line). Masses and external momenta
have the same values as for Fig.(\ref{fig6})}

\label{fig8}
\end{figure}

The branch point of the two-point function is on the real axes ($m_{12}^{2}>0$),
and hence the dispersion integral in Eq.(\ref{eq:32}) is well defined.
Since $B_{0}$ function is UV-divergent, the integral in Eq.(\ref{eq:32})
would diverge, which is addressed by introducing a cutoff $\Lambda$.
After differentiating numerically, $\Lambda$ dependence will cancel.
We checked numerically that the final result in Eq.(\ref{eq:32}) in
fact does not depend on the cutoff parameter $\Lambda$. Using LoopTools
libraries we can compare left and right hand parts of the Eq.(\ref{eq:32}),
and see good agreement (see Fig.(\ref{fig6})). 
In general, there will be cases when $m_{12}^{2}=m_{1}^{2}\bar{x}+m_{2}^{2}x-p_{1}^{2}x\bar{x}$ could become negative for the specific values of Feynman parameters or external
momenta. In this case, we would choose an integration contour over the upper half-part of the complex plane, and as a result perform dispersion integration from $-\Lambda^2$ to $\Lambda^2$ cut-off. To demonstrate, we consider a scalar two-point function with arbitrary imaginary mass. Since $B_0$ function is UV-divergent, we will take derivative of $B_0$ with respect to mass shift parameter $\lambda$, and for $\lim_{\lambda\rightarrow 0}\frac{\partial}{\partial\lambda}B_0(p^2,m_1^2,-|m_{2}^2|+\lambda)$ we can write:

\begin{align}
 & \lim_{\lambda\rightarrow 0}\frac{\partial}{\partial\lambda}B_0\left(p^2,m_1^2,-|m^2_{2}|+\lambda\right)=\frac{1}{2\pi i}\lim_{\lambda\rightarrow 0}\frac{\partial}{\partial\lambda}\intop_{-\Lambda^2}^{\Lambda^2}ds\frac{B_0\left(s,m_1^2,-|m_{2}^2|+\lambda\right) }{s-p^2-i\epsilon}\label{eq:32a}
\end{align}
If we take, for example, $\Lambda^2=10^{10}\ GeV^2$, $m_1^2=1\ GeV^2$ and $|m_2^2|=9\ GeV^2$, we observe that results obtained by means of Eq.(\ref{eq:32a}) and LoopTools shows up discrepancy only after 6th digit. 
\begin{table}
\begin{centering}
\begin{tabular}{|c|c|c|}
\hline 
$\, p^{2}$ $\left(\textrm{GeV}\right)^{2}$ & Eq.(\ref{eq:32a}) & LoopTools\tabularnewline
\hline 
\hline 
-5.0 & \,0.147987 - 0.079999 i \,& \,0.147988 - 0.079999 i \,\tabularnewline
\hline 
-1.0 & \,0.127835 - 0.037405 i  \,& \,0.127836 - 0.037405 i \,\tabularnewline
\hline 
-0.5 & \,0.124874 - 0.034242 i \,& \,0.124875 - 0.034242 i \,\tabularnewline
\hline 
0.5 & \,0.119143 - 0.028889 i  \,& \,0.119144 - 0.028889 i \,\tabularnewline
\hline 
1.0 & \,0.116402 - 0.026626 i \,& \,0.116403 - 0.026626 i \,\tabularnewline
\hline 
5.0 & \,0.097877 - 0.014968 i \,& \,0.097878 - 0.014968 i \,\tabularnewline
\hline 
10.0 & \,0.081819 - 0.008497 i \, & \,0.081820 - 0.008497 i \,\tabularnewline
\hline 
50.0 & \,0.039036 - 0.000911 i \,& \,0.039037 - 0.000911 i \,\tabularnewline
\hline 
\end{tabular}
\par\end{centering}
\caption{Comparison of the results for $\lim_{\lambda\rightarrow 0}\frac{\partial}{\partial\lambda}B_0(p^2,m_1^2,-|m_{2}^2|+\lambda)$ obtained in Eq.(\ref{eq:32a}) and LoopTools.}

\label{tbl2a}
\end{table}
Using the idea outlined in Eq.(\ref{eq:32}), we can also derive dispersive representation for
the higher order three-point function. With the help of Eq.(\ref{eq:26})
and Appendix A, we show in Figs.(\ref{fig7}) results for the $C_{1,2}$
and $C_{00,11,12,22}$ functions. 
Since $C_{00}$ is UV divergent, we obtained finite result after subtraction
at $q_{2}^{2}=0$. The results are in good agreement 
with LoopTools (see Fig.(\ref{fig8})). We also have tested third rank three
point functions and found an excellent agreement.
In general, we can write that triangle sub-loop can be replaced by
the following effective coupling
\begin{align}
 & \Gamma_{\Delta}=\hat{\mathbf{D}}\left[\frac{\Im F\left(s,m_{3}^{2},m_{12}^{2}+\lambda\right)}{s-\left(p_{2}+p_{1}\bar{x}\right)^{2}-i\epsilon}\right]\nonumber \\
\label{eq:33}\\
 & m_{12}^{2}=m_{1}^{2}\bar{x}+m_{2}^{2}x-p_{1}^{2}x\bar{x}.\nonumber 
\end{align}
Operator $\hat{\mathbf{D}}$ is defined as $\hat{\mathbf{D}}=\lim_{\lambda\rightarrow0}\frac{\partial}{\partial\lambda}\int_{0}^{1}dx\int_{r(x,\lambda)}^{\Lambda^{2}}ds...$, and function $r(x,\lambda)$ has  the following structure: $r(x,\lambda)=\left(m_{3}+\left(m_{12}^{2}+\lambda\right)^{1/2}\right)^{2}\theta(m_{12}^2)-\Lambda^2\theta(-m_{12}^2)$.
Momentum $p_{1}$ can be a combination of external momenta only. Momentum
$p_{2}$ can contain integration momentum of the second loop. The structure
of the function $\Im F(....)$ would depend on the nature of the particles
appearing in the triangle sub-loop and is specific to the process.
If subtractions are possible at the sub-loop level, then Eq.(\ref{eq:33})
has to be modified as follows:
\begin{align}
 & \hat{\Gamma}_{\Delta}=\hat{\mathbf{D}}\left[\frac{\Im F\left(s,m_{3}^{2},m_{12}^{2}+\lambda\right)\left[\left(p_{2}+p_{1}\bar{x}\right)^{2}-p_{1}^{2}\bar{x}^{2}\right]}{\left[s-\left(p_{2}+p_{1}\bar{x}\right)^{2}-i\epsilon\right]\left[s-p_{1}^{2}\bar{x}^{2}\right]}\right].\label{eq:34}
\end{align}
\begin{figure}
\begin{centering}
\includegraphics[scale=0.3]{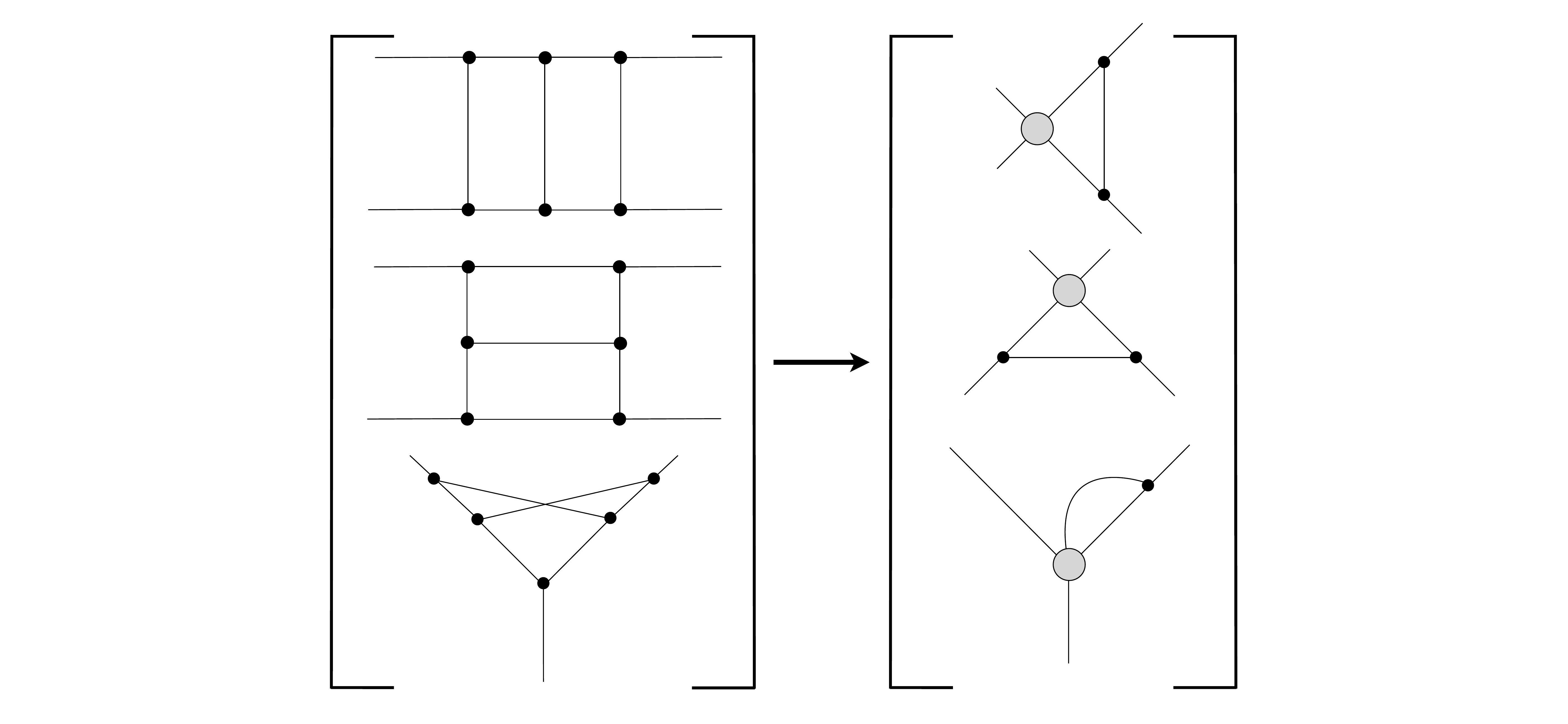}
\par\end{centering}
\caption{Box topology insertions in double boxes, ladder boxes and crossed-vertex graphs.}
\label{fig6a}
\end{figure}
The general algorithm of calculations with the triangle sub-loop insertions
could be summarized as follows. First, calculate one-loop triangle
insertion in Passarino-Veltman basis. Next, replace all three-point
functions by dispersive representation using rotation of the arguments,
so that momentum $p_{1}$ does not depend on the second-loop integration momentum. After
that, add the term $\left(s-\left(p_{2}+p_{1}\bar{x}\right)^{2}-i\epsilon\right)^{-1}$
from Eq.(\ref{eq:33}), or $\left[\left(p_{2}+p_{1}\bar{x}\right)^{2}-p_{1}^{2}\bar{x}^{2}\right]\left[s-\left(p_{2}+p_{1}\bar{x}\right)^{2}-i\epsilon\right]^{-1}\left[s-p_{1}^{2}\bar{x}^{2}\right]^{-1}$
from Eq.(\ref{eq:34}) to the second-loop integration. The next stage
is to get the second-loop integral, again in Passarino-Veltman basis,
and rewrite three-, four- and five-point functions in the two-point
basis. After that, apply the subtractions for the second loop using
counterterms in a given renormalization scheme. The final stage is
to evaluate the dispersion and Feynman parameters integrals, and do
numerical differentiation at least once.

\subsection{Box Sub-Loop}

Box sub-loops can be incorporated in double boxes, ladder boxes and crossed-vertex graphs (see Fig.(\ref{fig6a})).
As in the case of triangle sub-loop, our starting point would be an example of the double-box diagram. Here, we would define dispersive representation of the four-point insertion. In our case, all the external particles are on-shell, and masses on the top and bottom lines of the Fig.(\ref{fig8a}) are equal to $m_1$ and $m_2$, respectively. Following the momenta notation on Fig.(\ref{fig8a}), we can write for the left box sub-loop:
\begin{figure}
\begin{centering}
\includegraphics[scale=0.3]{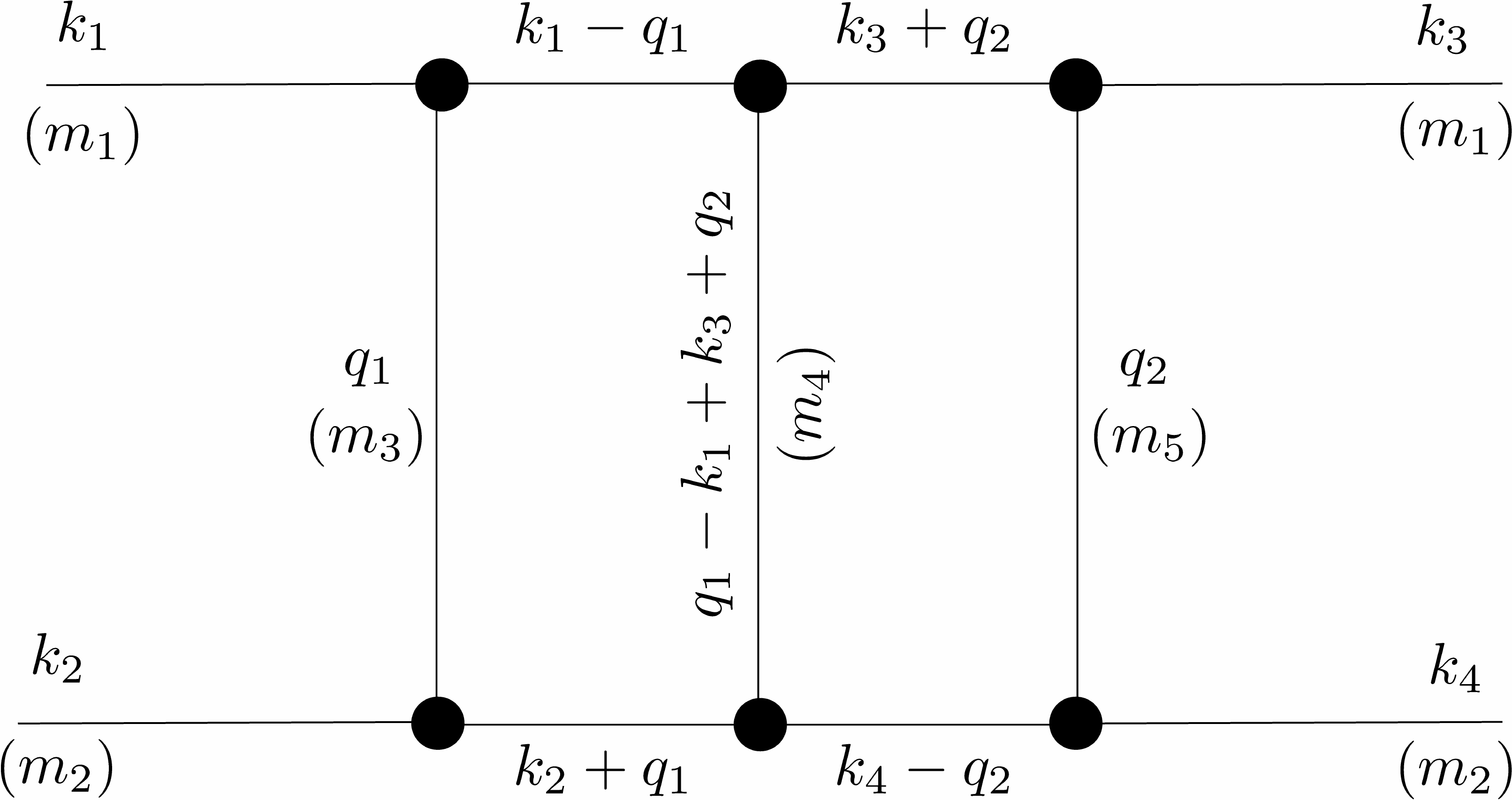}
\par\end{centering}
\caption{Double box diagram.}
\label{fig8a}
\end{figure}
\begin{align}
 & D_0=\frac{1}{i\pi^{2}}\int\frac{d^4 q_1}{\left[q_1^2-m_3^2\right]\left[\left(q_1+k_2\right)^2-m_1^2\right]\left[\left(q_1-k_1\right)^2-m_1^2\right]\left[\left(q_1+q_2+k_3-k_1\right)^2-m_4^2\right]}.\label{eq:35a}
\end{align}
After joining the first three propagators, using the mass-shift parameter approach, shifting momenta $\tau=q_1+q_2+k_3-k_1$ and applying dispersive representation of two-point function, we get the following result for the box sub-loop insertion:
\begin{align}
 D_0=\frac{1}{2 \pi i}\lim_{\lambda\rightarrow0}\frac{\partial^2}{\partial \lambda^2}\intop_0^1dx\intop_0^{1-x}dy\intop^{\Lambda^2}_{r\left(x,y,\lambda\right)}ds \Bigg[\frac{ 2i\,\Im B_0\left[s,m_4^2,m_{123}^2+\lambda\right]\theta\left(m_{123}^2\right)}{s-\left(q_2+k_3-xk_2-k_1\bar{y}\right)^2-i\epsilon}+\nonumber \\
\frac{B_0\left[s,m_4^2,m_{123}^2+\lambda\right]\theta\left(-m_{123}^2\right)}{s-\left(q_2+k_3-xk_2-k_1\bar{y}\right)^2-i\epsilon}\Bigg].
\label{eq:35b}
\end{align}
Mass $m^2_{123}=m_3^2\left(\bar{x}-y\right)+x^2m_2^2+y^2m_1^2-2xy\left(k_1k_2\right)$ could become negative for specific values of $x$ and $y$, so the function $r\left(x,y,\lambda\right)$ is defined as $r\left(x,y,\lambda\right)=\left(m_4+\left(m_{123}^2+\lambda\right)^{1/2}\right)^2\theta\left(m_{123}^2\right)-\Lambda^2\theta\left(-m_{123}^2\right)$. Here, $\theta \left(x\right)$ is a usual step function. The second-loop integral will receive, from Eq.(\ref{eq:35b}), an additional propagator and would become a four-point function with dispersion parameter $s$ playing role as the mass:
\begin{align}
 I_{d-box}=-\frac{1}{2\pi i}\hat{\mathbf{I}}_{\lambda xys} \left(2i\,\Im B_0\left[s,m_4^2,m_{123}^2+\lambda\right]\theta\left(m_{123}^2\right)+B_0\left[s,m_4^2,m_{123}^2+\lambda\right]\theta\left(-m_{123}^2\right)\right)\cdot \nonumber \\
\nonumber \\
\frac{1}{i\pi^2}\int \frac{d^4q_2}{\left[q_2^2-m_5^2\right]\left[\left(q_2+k_3\right)^2-m_1^2\right]\left[\left(q_2-k_4\right)^2-m_2^2\right]\left[\left(q_2+k_3-xk_2-k_1\bar{y}\right)^2-s-i\epsilon\right]}.
\label{eq:35bb}
\end{align}
Operator $\hat{\mathbf{I}}_{\lambda xys}$ has the following definition: $\hat{\mathbf{I}}_{\lambda xys}=\lim_{\lambda\rightarrow0}\frac{\partial^2}{\partial \lambda^2}\intop_0^1 dx\intop_0^{1-x}dy \intop_{r\left(x,y,\lambda\right)}^{\Lambda^2}ds...$. For the second-loop integration, we would also apply approach outlined in Eq.(\ref{eq:28}), and after joining the first three propagators, using mass-shift parameter $\xi$, final two-loop result can be expressed completely in two-point basis:
\begin{align}
 & I_{d-box}=&-\frac{1}{2\pi i}\hat{\mathbf{I}}_{\lambda xys}\hat{\mathbf{I}}_{\xi z\omega } \left(2i\Im B_0\left[s,m_4^2,m_{123}^2+\lambda\right]\theta\left(m_{123}^2\right)+B_0\left[s,m_4^2,m_{123}^2+\lambda\right] \theta\left(-m_{123}^2\right)\right)\cdot \nonumber \\
\nonumber \\
& &B_0\left[\left(\omega k_4+ \bar{z}k_3-x k_2-\bar{y}k_1\right)^2,m_{125}^2+\xi,s\right].\label{eq:35cc}
\end{align}
Operator $\hat{\mathbf{I}}_{\xi z\omega }$ is defined as $\hat{\mathbf{I}}_{\xi z\omega }=\lim_{\xi\rightarrow0}\frac{\partial^2}{\partial \xi^2}\intop_0^1 dz\intop_0^{1-z}d\omega...$, and mass $m_{125}^2=m_5^2\left(\bar{z}-\omega\right)+m_1^2z^2+\omega^2m_2^2-2z\omega \left(k_3k_4\right)$. For the generalized box sub-loop, we  can replace it by four-particle coupling 
\begin{align}
 & \Gamma_{\square}=\hat{\mathbf{D}}\left[\frac{\Im F\left(s,m_{4}^{2},m_{123}^{2}+\lambda\right)}{s-\left(p_{1}\left(\bar{x}-y\right)+p_{2}\bar{y}+p_{3}\right)^{2}-i\epsilon}\right]\nonumber \\
\label{eq:33aa}\\
 & m_{123}^{2}=m{}_{1}^{2}\left(\bar{x}-y\right)+m_{2}^{2}x+m_{3}^{2}y-p_{1}^{2}x\bar{x}-p_{12}^{2}y\bar{y}+2xy\left(p_{1}p_{12}\right),\nonumber \\
\nonumber 
\end{align}
where operator $\hat{\mathbf{D}}$ defined as $\hat{\mathbf{D}}=\lim_{\lambda\rightarrow0}\frac{\partial^2}{\partial\lambda^2}\int_{0}^{1}dx\intop^{1-x}_0dy\int_{r(x,y,\lambda)}^{\Lambda^2}ds...$. Using box sub-loop in Eq.(\ref{eq:33aa}), the second-loop integration will get an additional propagator and integration can be carried out in two-point function basis using Eqs.(\ref{eq:23}, \ref{eq:28}, \ref{eq:28a}).
\section{Numerical Example}
\begin{figure}
\begin{centering}
\includegraphics[scale=0.7]{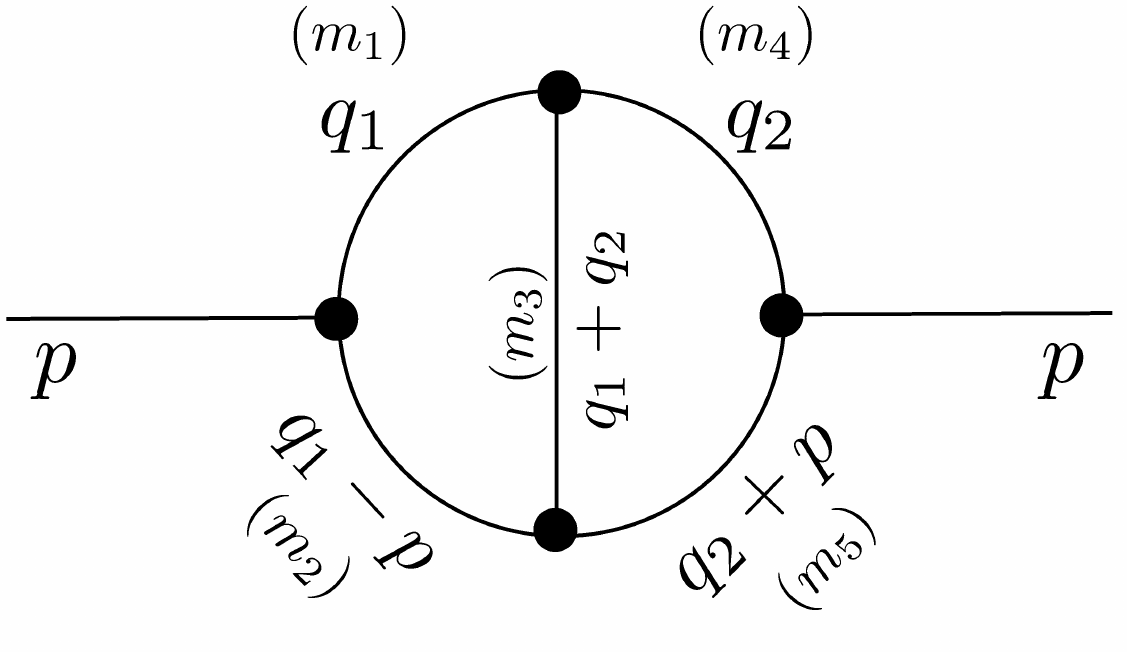}
\par\end{centering}
\caption{Examples where triangles topology insertion have been considered.}

\label{fig9}
\end{figure}
Lets consider a case with the triangle (three-point function) insertion applied as an example.
Namely, we will consider diagram on the Fig.(\ref{fig9}) with all
the masses are different and external particles are off-shell. Following the notation of \cite{Bohm},
a general expression for that graph is:
\begin{align}
 & I=-\frac{1}{\pi^{4}}\int\frac{d^{4}q_{1}d^{4}q_{2}}{\left[q_{1}^{2}-m_{1}^{2}\right]\left[\left(q_{1}-p\right)^{2}-m_{2}^{2}\right]\left[\left(q_{1}+q_{2}\right)^{2}-m_{3}^{2}\right]\left[q_{2}^{2}-m_{4}^{2}\right]\left[\left(q_{2}+p\right)^{2}-m_{5}^{2}\right]}.\label{eq:35}
\end{align}
After joining the first two propagators, shifting momenta as $\tau=q_{1}+q_{2}$,
and applying differentiation with mass shift parameter $\lambda$,
we get:
\begin{align}
 &I&=&\frac{1}{2\pi^{3}}\lim_{\lambda\rightarrow0}\frac{\partial}{\partial\lambda}\intop_{0}^{1}dx\intop_{r(x,\lambda)}^{\Lambda^{2}}ds\,\left(2i\Im B_{0}\left[s,m_{3}^{2},m_{12}^{2}+\lambda\right]\theta\left(m_{12}^2\right)+B_{0}\left[s,m_{3}^{2},m_{12}^{2}+\lambda\right]\theta\left(-m_{12}^2\right)\right)\cdot\nonumber \\
\nonumber\\
 &&&\int \frac{d^{4}q_{2}}{\left[q_{2}^{2}-m_{4}^{2}\right]\left[\left(q_{2}+xp\right)^{2}-s\right]\left[\left(q_{2}+p\right)^{2}-m_{5}^{2}\right]},\label{eq:36}
\end{align}
where $m_{12}^{2}=m_{1}^{2}\bar{x}+m_{2}^{2}x-p^{2}\bar{x}x$ and $r(x,\lambda)=\left(m_{3}+\left(m_{12}^{2}+\lambda\right)^{1/2}\right)^{2}\theta(m_{12}^2)-\Lambda^2\theta(-m_{12}^2)$. For
the time-like process when $p^{2}>0$, a value of $m_{12}^{2}$ could
become negative for the conditions above the threshold. That is adressed by using Eq.(\ref{eq:32a}). In space-like cases
of $p^{2}<0$, the dispersion representation of the insertion is well
defined. The second-loop integral, with the help of Eq.(\ref{eq:23}),
can be written in two-point function basis. The final two-loop result
can now be given in a compact form:
\begin{align}
I=-\frac{1}{2\pi i}\lim_{\left\{ \lambda,\xi\right\} \rightarrow0}\frac{\partial^{2}}{\partial\lambda\partial\xi}\intop_{0}^{1}dxdy\intop_{r(x,\lambda)}^{\Lambda^{2}}ds\,\Big(2i\Im B_{0}\left[s,m_{3}^{2},m_{12}^{2}+\lambda\right]\theta\left(m_{12}^2\right)+\nonumber \\
\nonumber \\
B_{0}\left[s,m_{3}^{2},m_{12}^{2}+\lambda\right]\theta\left(-m_{12}^2\right)\Big)B_{0}\left[p^{2}\left(x-y\right)^{2},s,m_{45}^{2}+\xi\right],\label{eq:37}
\end{align}
where $m_{45}^{2}=m_{4}^{2}\bar{y}+m_{5}^{2}y-p^{2}\bar{y}y$. The
integration and differentiation in Eq.(\ref{eq:37}) can be done numerically. Numerical integration, in the case of below threshold conditions, have used the global adaptive algorithm implemented in the integration package of Mathematica program. For the above threshold conditions, we have implemented numerical CUBA libraries from \cite{Hahn1, Hahn2}. These libraries are specifically designed for the multi-dimensional integration and employ Monte Carlo or cubature rules of polynomial degree algorithms. They showed the best convergence for the integration above threshold conditions. After that, we can compare our results with \cite{Bohm} (see Tbl.(\ref{tbl2})).
\begin{table}
\begin{centering}
\begin{tabular}{|c|c|c|c|c|}
\hline
$\,p^{2} \, \left(\textrm{GeV}\right)^{2}$&This work&$\Delta t_{\textrm{This Work}}$&\cite{Bohm}&$\Delta t_{\textrm{\cite{Bohm}}}$ \tabularnewline
\hline 
\hline
-50.0 & -0.08296 &75& - & - \tabularnewline
\hline 
-10.0 & -0.18399 &22& - & - \tabularnewline
\hline  
-5.0 & -0.22178 &17& - & - \tabularnewline
\hline 
-1.0 & -0.26919 &8& -&-\tabularnewline
\hline 
-0.5 &-0.27712 &9& -&-\tabularnewline
\hline 
-0.1 &-0.28360 &9& -&-\tabularnewline
\hline 
0.1 &-0.28714 &9& -0.28701&84\tabularnewline
\hline 
0.5 &-0.29443 &9& -0.29479&85\tabularnewline
\hline 
1.0 &-0.30449 &10& -0.30493&85 \tabularnewline
\hline 
5.0 & -0.45230&14 & -0.45241&86\tabularnewline
\hline 
10.0 &\, -0.48810 - 0.35318 i \,&30&\, -0.48825 - 0.35333 i  \,&86\tabularnewline
\hline
50.0 &\, 0.17335 - 0.11781 i \,&1120&\, 0.17391 - 0.11807 i  \,&85\tabularnewline
\hline
\end{tabular}
\par\end{centering}
\caption{Comparison of the results obtained in Eq.(\ref{eq:37}) with \cite{Bohm}.
We take masses the same as in \cite{Bohm}, $m_{1}^{2}=1$, $m_{2}^{2}=2$,
$m_{3}^{2}=3$, $m_{4}^{2}=4$ and $m_{5}^{2}=5$ $\textrm{GeV}^{2}$. Points $p^2>10$ $\textrm{GeV}^{2}$ correspond to above threshold condition. Third and fifth columns show computing time in seconds using Eq.(\ref{eq:37}) and \cite{Bohm} respectively} 

\label{tbl2}
\end{table}
We get a good agreement of the results and extend our calculations
to the space-like regime as well. In Tbl.(\ref{tbl2}), we also provide computing time and compare it to \cite{Bohm} for the case of two-dimensional integration using VEGAS routine. As we can see, computing time from \cite{Bohm} is nearly constant, but in our case it is highly dependent on the value of $\lvert p^2\rvert$. Further analysis shows that in our case we have to apply numerical differentiation with respect to mass-shift parameters, which is sensitive to the variations of $\lambda$ and $\xi$ in Eq.(\ref{eq:37}). Sensitivity grows when $\lvert p^2\rvert$ is getting large, and in order to achieve stability in numerical differentiation, precision of the integration over Feynman parameters should increase substantially. This is primarily the cause of the increase in integration time. 
The computing time of the dispersion integral is usually in the order of fraction of a second, and numerically very stable.
In the case of the box-type insertion, we would need, on top of the multidimensional integration, to deal with the second-order differentiation. That is a much more challenging task and will require two-loop graphs, with box insertions, to be evaluated using C++, Fortran or Python languages. This will be addressed in follow-up publications dedicated to the numerical evaluations of the dispersive representations of two-loop graphs with the box-type insertions.

\section{Conclusion}

In this work, we have applied the dispersive treatment approach of
the sub-loop insertion and represented the two-loop results in the
two-point function basis. The second-loop integration was reduced
to the two-point basis with the use of the partial tensor reduction.
The partial tensor reduction simplifies analytical expressions considerably
to the point that it is possible to employ computer algebra evaluating 
the two-loop calculations analytically and carry out integration and
differentiation numerically after that. As an example, we have compared
our results for the double self-energy shown on Fig.(\ref{fig9})
with \cite{Bohm} and found an excellent agreement.

\begin{acknowledgments}
The author is grateful to S. Barkanova, A. Czarnecki, A. Davydychev, V. Pascalutsa, H. Patel, H. Spiesberger and M. Vanderhaeghen for the fruitful and exciting discussions. The author would also like
to express a special thanks to the Mainz Institute for Theoretical
Physics (MITP) for its hospitality and support. This work was supported
by the Natural Sciences and Engineering Research Council (NSERC) of
Canada.
\end{acknowledgments}

\section{Appendix}

\subsection{$C_{\mu\nu}$ and $C_{\mu\nu\alpha}$}

With the help of operator $\hat{\mathbf{I}}_{C}=\lim_{\lambda\rightarrow0}\frac{\partial}{\partial\lambda}\intop_{0}^{1}dx...$,
we can write

$C_{\mu\nu}$:

\begin{align*}
 & C_{00}=\hat{\mathbf{I}}_{C}\left[B_{00}\right] &  & C_{12}=-\hat{\mathbf{I}}_{C}\left[\left(B_{1}+B_{11}\right)x\right]\\
 & C_{11}=\hat{\mathbf{I}}_{C}\left[B_{11}x^{2}\right] &  & C_{22}=\hat{\mathbf{I}}_{C}\left[B_{0}+2B_{1}+B_{11}\right].
\end{align*}

$C_{\mu\nu\alpha}$:
\begin{align*}
 & C_{001}=\hat{\mathbf{I}}_{C}\left[B_{001}x\right] &  & C_{112}=-\hat{\mathbf{I}}_{C}\left[\left(B_{11}+B_{111}\right)x^{2}\right]\\
 & C_{002}=-\hat{\mathbf{I}}_{C}\left[B_{00}+B_{001}\right] &  & C_{122}=\hat{\mathbf{I}}_{C}\left[\left(B_{1}+2B_{11}+B_{111}\right)x\right]\\
 & C_{111}=\hat{\mathbf{I}}_{C}\left[B_{111}x^{3}\right] &  & C_{222}=-\hat{\mathbf{I}}_{C}\left[B_{0}+3\left(B_{1}+B_{11}\right)+B_{111}\right].
\end{align*}
Two point functions $B_{i,ij,ijk}$ have the following definition:
$B_{i,ij,ijk}\equiv B_{i,ij,ijk}\left[\left(p_{1}\bar{x}+p_{2}\right)^{2},m_{3}^{2},m_{12}^{2}+\lambda\right]$.

\subsection{$D_{\mu}$, $D_{\mu\nu}$, $D_{\mu\nu\rho}$ and $D_{\mu\nu\rho\sigma}$}

Introducing operator $\hat{\mathbf{I}}_{D}=\lim_{\lambda\rightarrow0}\frac{\partial^{2}}{\partial\lambda^{2}}\int_{0}^{1}dx\int_{0}^{1-x}dy...$,
we list only the final results

$D_{\mu}:$
\begin{align*}
 & D_{1}=\hat{\mathbf{I}}_{D}\left[B_{1}x\right]\\
 & D_{2}=\hat{\mathbf{I}}_{D}\left[B_{1}y\right]\\
 & D_{3}=-\hat{\mathbf{I}}_{D}\left[B_{0}+B_{1}\right].
\end{align*}

$D_{\mu\nu}$:
\begin{align*}
 & D_{00}=\hat{\mathbf{I}}_{D}\left[B_{00}\right] &  & D_{33}=\hat{\mathbf{I}}_{D}\left[B_{0}+2B_{1}+B_{11}\right]\\
 & D_{11}=\hat{\mathbf{I}}_{D}\left[B_{11}x^{2}\right] &  & D_{12}=\hat{\mathbf{I}}_{D}\left[B_{11}xy\right]\\
 & D_{22}=\hat{\mathbf{I}}_{D}\left[B_{11}y^{2}\right] &  & D_{13}=-\hat{\mathbf{I}}_{D}\left[\left(B_{1}+B_{11}\right)x\right]\\
 &  &  & D_{23}=-\hat{\mathbf{I}}_{D}\left[\left(B_{1}+B_{11}\right)y\right].
\end{align*}

$D_{\mu\nu\rho}$:
\begin{align*}
 & D_{001}=\hat{\mathbf{I}}_{D}\left[B_{001}x\right] &  & D_{122}=\hat{\mathbf{I}}_{D}\left[B_{111}xy^{2}\right]\\
 & D_{002}=\hat{\mathbf{I}}_{D}\left[B_{001}y\right] &  & D_{123}=-\hat{\mathbf{I}}_{D}\left[\left(B_{11}+B_{111}\right)xy\right]\\
 & D_{003}=-\hat{\mathbf{I}}_{D}\left[B_{00}+B_{001}\right] &  & D_{222}=\hat{\mathbf{I}}_{D}\left[B_{111}y^{3}\right]\\
 & D_{111}=\hat{\mathbf{I}}_{D}\left[B_{111}x^{3}\right] &  & D_{223}=-\hat{\mathbf{I}}_{D}\left[\left(B_{11}+B_{111}\right)y^{2}\right]\\
 & D_{112}=\hat{\mathbf{I}}_{D}\left[B_{111}x^{2}y\right] &  & D_{233}=\hat{\mathbf{I}}_{D}\left[\left(B_{1}+2B_{11}+B_{111}\right)y\right]\\
 & D_{113}=-\hat{\mathbf{I}}_{D}\left[\left(B_{11}+B_{111}\right)x^{2}\right] &  & D_{333}=-\hat{\mathbf{I}}_{D}\left[B_{0}+3\left(B_{1}+B_{11}\right)+B_{111}\right].
\end{align*}

$D_{\mu\nu\rho\sigma}$:
\begin{align*}
 & D_{0000}=\hat{\mathbf{I}}_{D}\left[B_{0000}\right] &  & D_{1123}=-\hat{\mathbf{I}}_{D}\left[\left(B_{111}+B_{1111}\right)x^{2}y\right]\\
 & D_{0011}=\hat{\mathbf{I}}_{D}\left[B_{0011}x^{2}\right] &  & D_{1133}=\hat{\mathbf{I}}_{D}\left[\left(B_{11}+2B_{111}+B_{1111}\right)x^{2}\right]\\
 & D_{0012}=\hat{\mathbf{I}}_{D}\left[B_{0011}xy\right] &  & D_{1222}=\hat{\mathbf{I}}_{D}\left[B_{1111}xy^{3}\right]\\
 & D_{0013}=-\hat{\mathbf{I}}_{D}\left[\left(B_{001}+B_{1111}\right)x\right] &  & D_{1223}=-\hat{\mathbf{I}}_{D}\left[\left(B_{111}+B_{1111}\right)xy^{2}\right]\\
 & D_{0022}=\hat{\mathbf{I}}_{D}\left[B_{0011}x^{2}\right] &  & D_{1233}=\hat{\mathbf{I}}_{D}\left[\left(B_{11}+2B_{111}+B_{1111}\right)xy\right]\\
 & D_{0023}=-\hat{\mathbf{I}}_{D}\left[\left(B_{001}+B_{1111}\right)y\right] &  & D_{1333}=-\hat{\mathbf{I}}_{D}\left[\left(B_{1}+3\left(B_{11}+B_{111}\right)+B_{1111}\right)x\right]\\
 & D_{0033}=\hat{\mathbf{I}}_{D}\left[B_{00}+2B_{001}+B_{0011}\right] &  & D_{2222}=\hat{\mathbf{I}}_{D}\left[B_{1111}y^{4}\right]\\
 & D_{1111}=\hat{\mathbf{I}}_{D}\left[B_{1111}x^{4}\right] &  & D_{2223}=-\hat{\mathbf{I}}_{D}\left[\left(B_{111}+B_{1111}\right)y^{3}\right]\\
 & D_{1112}=\hat{\mathbf{I}}_{D}\left[B_{1111}x^{3}y\right] &  & D_{2233}=\hat{\mathbf{I}}_{D}\left[\left(B_{11}+2B_{111}+B_{1111}\right)y^{2}\right]\\
 & D_{1113}=-\hat{\mathbf{I}}_{D}\left[\left(B_{111}+B_{1111}\right)x^{3}\right] &  & D_{2333}=-\hat{\mathbf{I}}_{D}\left[\left(B_{1}+3\left(B_{11}+B_{111}\right)+B_{1111}\right)y\right]\\
 & D_{1122}=\hat{\mathbf{I}}_{D}\left[B_{1111}x^{2}y^{2}\right] &  & D_{3333}=\hat{\mathbf{I}}_{D}\left[\left(B_{0}+4\left(B_{1}+B_{111}\right)+6B_{11}+B_{1111}\right)\right].
\end{align*}
Two point functions $B_{i,ij,ijk,ijkl}$ have the following arguments:
$B_{i,ij,ijk,ijkl}\equiv B_{i,ij,ijk,ijkl}\left[\left(p_{1}\left(\bar{x}-y\right)+p_{2}\bar{y}+p_{3}\right)^{2},m_{4}^{2},m_{123}^{2}+\lambda\right]$.
As it can be seen reduction of $D_{ijkl}$ uses $B_{0000}$, $B_{0011}$
and $B_{1111}$ functions, which can be evaluated using \cite{Denner}
including the cases of imaginary masses. 

\subsection{$E_{\mu}$, $E_{\mu\nu}$, $E_{\mu\nu\rho}$ and $E_{\mu\nu\rho\sigma}$}

Using the operator $\hat{\mathbf{I}}_{E}=\lim_{\lambda\rightarrow0}\frac{\partial^{3}}{\partial\lambda^{3}}\int_{0}^{1}dx\int_{0}^{1-x}dy\int_{0}^{1-x-y}dz...$,
we list only the final results

$E_{\mu}:$
\begin{align*}
 & E_{1}=\hat{\mathbf{I}}_{E}\left[B_{1}x\right]\\
 & E_{2}=\hat{\mathbf{I}}_{E}\left[B_{1}y\right]\\
 & E_{3}=\hat{\mathbf{I}}_{E}\left[B_{1}z\right]\\
 & E_{3}=-\hat{\mathbf{I}}_{E}\left[B_{0}+B_{1}\right].
\end{align*}

$E_{\mu\nu}$:
\begin{align*}
 & E_{00}=\hat{\mathbf{I}}_{E}\left[B_{00}\right] &  & E_{23}=\hat{\mathbf{I}}_{E}\left[B_{11}yz\right]\\
 & E_{11}=\hat{\mathbf{I}}_{E}\left[B_{11}x^{2}\right] &  & E_{24}=-\hat{\mathbf{I}}_{E}\left[\left(B_{1}+B_{11}\right)y\right]\\
 & E_{12}=\hat{\mathbf{I}}_{E}\left[B_{11}xy\right] &  & E_{33}=\hat{\mathbf{I}}_{E}\left[B_{11}z^{2}\right]\\
 & E_{13}=\hat{\mathbf{I}}_{E}\left[B_{11}xz\right] &  & E_{34}=-\hat{\mathbf{I}}_{E}\left[\left(B_{1}+B_{11}\right)z\right]\\
 & E_{14}=-\hat{\mathbf{I}}_{E}\left[\left(B_{1}+B_{11}\right)x\right] &  & E_{44}=\hat{\mathbf{I}}_{E}\left[B_{0}+2B_{1}+B_{11}\right].\\
 & E_{22}=\hat{\mathbf{I}}_{E}\left[B_{11}y^{2}\right]
\end{align*}

$E_{\mu\nu\rho}$:
\begin{align*}
 & E_{001}=\hat{\mathbf{I}}_{E}\left[B_{001}x\right] &  & E_{134}=-\hat{\mathbf{I}}_{E}\left[\left(B_{11}+B_{111}\right)xz\right]\\
 & E_{002}=\hat{\mathbf{I}}_{E}\left[B_{001}y\right] &  & E_{144}=\hat{\mathbf{I}}_{E}\left[\left(B_{1}+2B_{11}+B_{111}\right)x\right]\\
 & E_{003}=\hat{\mathbf{I}}_{E}\left[B_{001}z\right] &  & E_{222}=\hat{\mathbf{I}}_{E}\left[B_{111}y^{3}\right]\\
 & E_{004}=-\hat{\mathbf{I}}_{E}\left[B_{00}+B_{001}\right] &  & E_{223}=\hat{\mathbf{I}}_{E}\left[B_{111}y^{2}z\right]\\
 & E_{111}=\hat{\mathbf{I}}_{E}\left[B_{111}x^{3}\right] &  & E_{224}=-\hat{\mathbf{I}}_{E}\left[\left(B_{11}+B_{111}\right)y^{2}\right]\\
 & E_{112}=\hat{\mathbf{I}}_{E}\left[B_{111}x^{2}y\right] &  & E_{233}=\hat{\mathbf{I}}_{E}\left[B_{111}yz^{2}\right]\\
 & E_{113}=\hat{\mathbf{I}}_{E}\left[B_{111}x^{2}z\right] &  & E_{234}=-\hat{\mathbf{I}}_{E}\left[\left(B_{11}+B_{111}\right)yz\right]\\
 & E_{114}=-\hat{\mathbf{I}}_{E}\left[\left(B_{11}+B_{111}\right)x^{2}\right] &  & E_{244}=\hat{\mathbf{I}}_{E}\left[\left(B_{1}+2B_{11}+B_{111}\right)y\right]\\
 & E_{122}=\hat{\mathbf{I}}_{E}\left[B_{111}xy^{2}\right] &  & E_{333}=\hat{\mathbf{I}}_{E}\left[B_{111}z^{3}\right]\\
 & E_{123}=\hat{\mathbf{I}}_{E}\left[B_{111}xyz\right] &  & E_{334}=-\hat{\mathbf{I}}_{E}\left[\left(B_{11}+B_{111}\right)z^{2}\right]\\
 & E_{124}=-\hat{\mathbf{I}}_{E}\left[\left(B_{11}+B_{111}\right)xy\right] &  & E_{344}=\hat{\mathbf{I}}_{E}\left[\left(B_{1}+2B_{11}+B_{111}\right)z\right]\\
 & E_{133}=\hat{\mathbf{I}}_{E}\left[B_{111}xz^{2}\right] &  & E_{444}=-\hat{\mathbf{I}}_{E}\left[B_{0}+3\left(B_{1}+B_{11}\right)+B_{111}\right].
\end{align*}

$E_{\mu\nu\rho\sigma}$:
\begin{align*}
 & E_{0000}=\hat{\mathbf{I}}_{E}\left[B_{0000}\right] &  & E_{1224}=-\hat{\mathbf{I}}_{E}\left[\left(B_{111}+B_{1111}\right)xy^{2}\right]\\
 & E_{0011}=\hat{\mathbf{I}}_{E}\left[B_{0011}x^{2}\right] &  & E_{1233}=\hat{\mathbf{I}}_{E}\left[B_{1111}xyz^{2}\right]\\
 & E_{0012}=\hat{\mathbf{I}}_{E}\left[B_{0011}xy\right] &  & E_{1234}=-\hat{\mathbf{I}}_{E}\left[\left(B_{111}+B_{1111}\right)xyz\right]\\
 & E_{0013}=\hat{\mathbf{I}}_{E}\left[B_{0011}xz\right] &  & E_{1244}=\hat{\mathbf{I}}_{E}\left[\left(B_{11}+2B_{111}+B_{1111}\right)xy\right]\\
 & E_{0014}=-\hat{\mathbf{I}}_{E}\left[\left(B_{001}+B_{1111}\right)x\right] &  & E_{1333}=\hat{\mathbf{I}}_{E}\left[B_{1111}xz^{3}\right]\\
 & E_{0022}=\hat{\mathbf{I}}_{E}\left[B_{0011}y^{2}\right] &  & E_{1334}=-\hat{\mathbf{I}}_{E}\left[\left(B_{111}+B_{1111}\right)xz^{3}\right]\\
 & E_{0023}=\hat{\mathbf{I}}_{E}\left[B_{0011}yz\right] &  & E_{1344}=\hat{\mathbf{I}}_{E}\left[\left(B_{11}+2B_{111}+B_{1111}\right)xz\right]\\
 & E_{0024}=-\hat{\mathbf{I}}_{E}\left[\left(B_{001}+B_{1111}\right)y\right] &  & E_{1444}=-\hat{\mathbf{I}}_{E}\left[\left(B_{1}+3\left(B_{11}+B_{111}\right)+B_{1111}\right)x\right]\\
 & E_{0033}=\hat{\mathbf{I}}_{E}\left[B_{0011}z^{2}\right] &  & E_{2222}=\hat{\mathbf{I}}_{E}\left[B_{1111}y^{4}\right]\\
 & E_{0034}=-\hat{\mathbf{I}}_{E}\left[\left(B_{001}+B_{1111}\right)z\right] &  & E_{2223}=\hat{\mathbf{I}}_{E}\left[B_{1111}y^{2}z\right]\\
 & E_{0044}=\hat{\mathbf{I}}_{E}\left[B_{00}+2B_{001}+B_{0011}\right] &  & E_{2224}=-\hat{\mathbf{I}}_{E}\left[\left(B_{111}+B_{1111}\right)y^{3}\right]\\
 & E_{1111}=\hat{\mathbf{I}}_{E}\left[B_{1111}x^{4}\right] &  & E_{2233}=\hat{\mathbf{I}}_{E}\left[B_{1111}y^{2}z^{2}\right]\\
 & E_{1112}=\hat{\mathbf{I}}_{E}\left[B_{1111}x^{3}y\right] &  & E_{2234}=-\hat{\mathbf{I}}_{E}\left[\left(B_{111}+B_{1111}\right)y^{2}z\right]\\
 & E_{1113}=\hat{\mathbf{I}}_{D}\left[B_{1111}x^{3}z\right] &  & E_{2244}=\hat{\mathbf{I}}_{E}\left[\left(B_{11}+2B_{111}+B_{1111}\right)y^{2}\right]\\
 & E_{1114}=-\hat{\mathbf{I}}_{E}\left[\left(B_{111}+B_{1111}\right)x^{3}\right] &  & E_{2333}=\hat{\mathbf{I}}_{E}\left[B_{1111}yz^{3}\right]\\
 & E_{1122}=\hat{\mathbf{I}}_{E}\left[B_{1111}x^{2}y^{2}\right] &  & E_{2334}=-\hat{\mathbf{I}}_{E}\left[\left(B_{111}+B_{1111}\right)yz^{2}\right]\\
 & E_{1123}=\hat{\mathbf{I}}_{E}\left[B_{1111}x^{2}yz\right] &  & E_{2344}=\hat{\mathbf{I}}_{E}\left[\left(B_{11}+2B_{111}+B_{1111}\right)yz\right]\\
 & E_{1124}=-\hat{\mathbf{I}}_{E}\left[\left(B_{111}+B_{1111}\right)x^{2}y\right] &  & E_{2444}=-\hat{\mathbf{I}}_{E}\left[\left(B_{1}+3\left(B_{11}+B_{111}\right)+B_{1111}\right)y\right]\\
 & E_{1133}=\hat{\mathbf{I}}_{E}\left[B_{1111}x^{2}z^{2}\right] &  & E_{3333}=\hat{\mathbf{I}}_{E}\left[B_{1111}z^{4}\right]\\
 & E_{1134}=-\hat{\mathbf{I}}_{E}\left[\left(B_{111}+B_{1111}\right)x^{2}z\right] &  & E_{3334}=-\hat{\mathbf{I}}_{E}\left[\left(B_{111}+B_{1111}\right)z^{3}\right]\\
 & E_{1144}=\hat{\mathbf{I}}_{E}\left[\left(B_{11}+2B_{111}+B_{1111}\right)x^{2}\right] &  & E_{3344}=\hat{\mathbf{I}}_{E}\left[\left(B_{11}+2B_{111}+B_{1111}\right)z^{2}\right]\\
 & E_{1222}=\hat{\mathbf{I}}_{E}\left[B_{1111}xy^{3}\right] &  & E_{3444}=-\hat{\mathbf{I}}_{E}\left[\left(B_{1}+3\left(B_{11}+B_{111}\right)+B_{1111}\right)z\right]\\
 & E_{1223}=\hat{\mathbf{I}}_{E}\left[B_{1111}xy^{2}z\right] &  & E_{4444}=\hat{\mathbf{I}}_{E}\left[\left(B_{0}+4\left(B_{1}+B_{111}\right)+6B_{11}+B_{1111}\right)\right].
\end{align*}
Two point functions $B_{i,ij,ijk,ijkl}$ are defined as follows:
$B_{i,ij,ijk,ijkl}\equiv B_{i,ij,ijk,ijkl}\left[\left(p_{1}\left(\bar{x}-y-z\right)+p_{2}\left(\bar{y}-z\right)+p_{3}\bar{z}+p_{4}\right)^{2},m_{5}^{2},m_{1234}^{2}+\lambda\right]$.

\end{document}